\newcount\draft\draft=0		
\newcount\cameraready\cameraready=1	

\ifnum\cameraready=0
 \documentclass[10pt]{sig-alternate-05-2015}
\else
 \documentclass[10pt]{sig-alternate-10pt}
\fi

\usepackage{graphicx,epsfig,sped,times}
\usepackage{algorithm}
\usepackage{units}
\usepackage{clrscode}
\usepackage[noeka]{mathrmletter}
\usepackage{subfigure}
\usepackage{gensymb}
\usepackage{ulem}
\renewcommand{\em}{\normalem}
\usepackage{multirow}
\usepackage[noend]{algorithmic}
\usepackage{epstopdf}
\usepackage{xspace}
\usepackage{mdwlist}
\usepackage{balance}
\usepackage{url}
\usepackage[breaklinks]{hyperref}
\usepackage{breakurl}

\usepackage{cite}

\newfont{\mycrnotice}{ptmr8t at 7pt}
\newfont{\myconfname}{ptmri8t at 7pt}
\let\crnotice\mycrnotice%
\let\confname\myconfname%

\newcommand{\name} {interscatter} 
\newcommand{\Name} {Interscatter} 
\newfont{\coprimary}{phvr8t at 10pt}

\usepackage{color, colortbl}
\usepackage{graphicx}
\DeclareGraphicsExtensions{.pdf,.jpg,.png}
\graphicspath{{.}{./figures/}{./results/}}

\usepackage{listings}
\newenvironment{Itemize}%
{\begin{itemize}%
\setlength{\itemindent}{0em}
\setlength{\itemsep}{0pt}%
\setlength{\topsep}{0pt}%
\setlength{\parsep}{-2pt}
\setlength{\partopsep}{0pt}%
\setlength{\parskip}{0pt}}%
{\end{itemize}}
{\begin{enumerate}%
\setlength{\itemsep}{0pt}%
\setlength{\topsep}{0pt}%
\setlength{\partopsep}{0pt}%
\setlength{\parskip}{0pt}}%
{\end{enumerate}}
\makeatletter
  \newcommand\figcaption{\def\@captype{figure}\caption}
  \newcommand\tabcaption{\def\@captype{table}\caption}
\makeatother

\newcommand{\xqed}{\nobreak \ifvmode \relax \else
      \ifdim\lastskip<1.5em \hskip-\lastskip
      \hskip1.5em plus0em minus0.5em \fi \nobreak
      \vrule height0.75em width0.5em depth0.25em\fi}

\newcommand{\xref}[1]{\S\ref{#1}}

\newcommand{\textred}[1]{\textcolor{black}{#1}}
\ifx\noeditingmarks\undefined
   \newcommand{\pgwrapper}[2]{\textred{#1: #2}}
\else
   \newcommand{\pgwrapper}[2]{}
\fi

\usepackage{xxx} 

\begin{document}


\font\ttlfnt=phvb8t at 16pt

\newcommand{\supsym}[1]{\raisebox{6pt}{{\footnotesize #1}}}

\widowpenalty = 10000

\title{\textred{Inter-Technology Backscatter: Towards \\Internet Connectivity for Implanted Devices}}

\numberofauthors{1}
\author{%
Vikram Iyer\supsym{$\dagger$}, Vamsi Talla\supsym{$\dagger$}, Bryce Kellogg\supsym{$\dagger$}, Shyamnath Gollakota and Joshua R. Smith\\
\affaddr{University of Washington}\\
\coprimary{\supsym{$\dagger$}Co-primary Student Authors}\\
\affaddr{\{vsiyer, vamsit, kellogg, gshyam, jrsjrs\}@uw.edu}\\
}

\maketitle

\begin{sloppypar}

{\bf Abstract --} \textred{We introduce inter-technology backscatter, a novel approach that transforms wireless transmissions from one technology to another, on the air. Specifically, we show for the first time that Bluetooth transmissions can be used to create Wi-Fi and ZigBee-compatible signals using backscatter communication. Since Bluetooth, Wi-Fi and ZigBee radios are widely available, this approach enables a backscatter design that works using only commodity devices.}

We build prototype backscatter hardware using an FPGA and experiment with various Wi-Fi, Bluetooth and ZigBee devices. Our experiments show we can create 2--11~Mbps Wi-Fi standards-compliant signals by backscattering Bluetooth transmissions. To show the generality of our approach, we also demonstrate generation of standards-complaint ZigBee signals by backscattering Bluetooth transmissions. Finally, we build proof-of-concepts for previously infeasible applications including the first contact lens form-factor antenna prototype and an implantable neural recording interface that communicate directly with commodity devices such as smartphones and watches, \textred{ thus enabling the vision of Internet connected implanted devices. }

%
%
%

\section{Introduction}
There has been recent interest in medical applications including smart contact lens platforms~\cite{lens1,lens2,lens3,lens4} that measure biomarkers like glucose, cholesterol and sodium for diabetes management, as well as implantable neural devices that help in the treatment of epilepsy, Parkinson's disease~\cite{ecog-parkinson}, reanimation of limbs~\cite{ecog-limb} and development of brain-computer interfaces~\cite{ecog-bmi}. As part of an ecosystem of connected devices, these implants have the potential to transform the management of chronic medical conditions and  enable novel interaction capabilities.

 In this paper we ask the following question: can these implanted devices communicate directly with mobile devices such as smartphones, watches and tablets?  The key challenge is that owing to their severe power constraints, these devices cannot use conventional radios to generate Wi-Fi, Bluetooth or ZigBee transmissions and hence cannot communicate directly with commodity devices.  While recent work on passive Wi-Fi~\cite{nsdi16} significantly reduces the power consumption of Wi-Fi transmissions using backscatter communication, it requires infrastructure support in the form of specialized hardware that can generate the continuous wave RF signal needed for backscattering Wi-Fi signals.  Thus, it cannot fully plug and play with existing mobile devices.

\begin{figure}[!t]
        \includegraphics[width=\columnwidth]{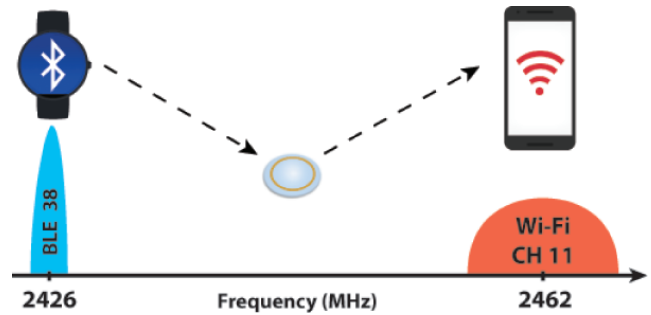}
  \caption{{\bf \Name\ Communication.} The backscatter device (e.g., contact lens) reflects transmissions from a Bluetooth device (e.g., smart watch) to generate standards-compatible Wi-Fi signals that are decodable on a Wi-Fi device (e.g., smartphone).}
\label{fig:page1}
\vskip -0.15in
\end{figure}

\begin{figure*}[!t]
\centering
        \includegraphics[width=0.8\textwidth]{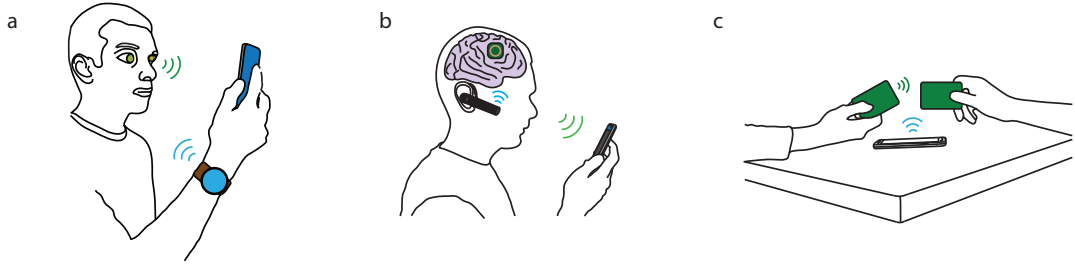}
\vskip -0.1in
  \caption{{\bf Potential applications of \Name\ communication.}
(a) Active contact lens systems can backscatter Bluetooth transmissions from a watch to generate Wi-Fi signals to a phone, (b) implantable brain interfaces can communicate by using Bluetooth headsets and smartphones, and (c) credit cards can communicate with each other by backscattering Bluetooth transmissions from a smartphone.}
\label{fig:applications}
\vskip -0.15in
\end{figure*}

We introduce {\it\name}\footnote{Short for inter-technology backscatter communication.}, a novel backscatter communication system that works using only commodity devices by transforming transmissions from one technology to the other, on the air. We observe that most mobile devices have a Bluetooth, Wi-Fi or ZigBee radio. Further, users increasingly carry smart watches, fitness trackers and Bluetooth headsets in addition to their smartphones. \Name\ reuses the radios on these devices as both RF sources for backscatter as well as receivers the backscattered signals. This allows us to benefit from the Wi-Fi, ZigBee and Bluetooth economies of scale (a few dollars per chip~\cite{ti_ble_chip_buy}) and eliminate the cost of a dedicated  reader. Further, it significantly lowers the barrier to adoption, as the wireless hardware is widely available on commodity devices and does not require the user to carry a specialized backscatter reader.

Our key idea is a novel approach to backscatter communication: create Wi-Fi signals by backscattering Bluetooth transmissions. To understand this, consider an \name\ device in the presence of a Bluetooth transmitter (e.g., smart watch) and a Wi-Fi receiver (e.g., smartphone), as shown in Fig.~\ref{fig:page1}. We backscatter advertising packets from the Bluetooth device to synthesize 802.11b signals at 2--11~Mbps. In the figure, the \name\ device backscatters Bluetooth packets on channel 38 to generate standards-compliant 802.11b packets on Wi-Fi channel 11.
Realizing this idea is challenging for at least three reasons:
\begin{Itemize}
\item Wi-Fi and Bluetooth have different physical layer specifications ---  Wi-Fi requires a 22~MHz bandwidth and uses spread spectrum coding. Bluetooth occupies a 1-2~MHz bandwidth and uses Gaussian Frequency Shift Keying (GFSK).

\item Bluetooth operates at carrier frequencies that are different from Wi-Fi. While sideband-backscatter modulation~\cite{nsdi16,bluetoothbackscatter} could shift the carrier by tens of Megahertz, it also creates a redundant copy on the opposite side of the carrier.\footnote{Say an RF source transmits a signal $\sin(f_ct)$, sideband modulation backscatters at a frequency of $\Delta f$, resulting in $sin(\Delta f t)$,  which would in turn create the multiplicative signal $2\sin(f_ct)\sin(\Delta ft) = \cos((f_c-\Delta f)t) - \cos((f_c+\Delta f)t)$. The two signals correspond to a signal with a positive frequency shift and its mirror copy with a negative shift.} This not only wastes bandwidth but the redundant copy would also lie outside the unlicensed ISM band (see~\xref{sec:singleside}).

\item For  bi-directional communication, we need a receiver at the \name\ device. Wi-Fi and Bluetooth receivers consume orders of magnitude higher power than backscatter and would offset its power saving. In fact, existing ultra-low power receiver designs rely on amplitude modulation (AM)~\cite{haykin2008communication}, which is not supported by Wi-Fi or Bluetooth.


\end{Itemize}
At a high level, we first transform a Bluetooth transmission into a single tone signal and use backscatter to create standards-compliant Wi-Fi packets on a single side of the resulting single tone Bluetooth signal. Specifically, we make three key technical contributions to achieve this design.

\begin{Itemize}
\item  We show for the first time that {\it Bluetooth radios  can be used to create single-tone transmissions}. We leverage that Bluetooth uses GFSK that encodes bits using two frequency tones. Thus, if we could transmit a stream of constant ones or zeros, we can create a single tone transmission. In~ \xref{sec:bluetooth}, we describe how to achieve this on commodity Bluetooth devices in the presence of the data whitening, CRC and headers.


\item We present the {\it first single-sideband backscatter design} that creates frequency shifts on a single side of the carrier. This lets us create 2--11~Mbps Wi-Fi signals shifted by tens of Megahertz on only {\it one side} of our single tone Bluetooth transmissions. We achieve this using complex impedances at the backscatter switch, without the need for a power-consuming 2.4~GHz oscillators (see~\xref{sec:genwifi}).

\item  We {\it transform OFDM Wi-Fi devices into AM modulators}. At a high level, the Wi-Fi device in our design modulates the amplitude profile of OFDM symbols to create an AM signal. We show that this can be achieved by just setting the appropriate data bits in a Wi-Fi packet, without the need for any hardware power control.~\xref{sec:downlink} describes how we perform this in the presence of Wi-Fi scrambling, convolutional encoding and interleaving.

\end{Itemize}

We build prototype backscatter hardware on an FPGA platform and experiment with various Bluetooth and Wi-Fi devices. Our evaluation shows that we can generate 2--11~Mbps Wi-Fi signals from Bluetooth transmissions. To estimate the power consumption, we also design an integrated circuit using Cadence and Synopsis~\cite{cadence_RFspectre, synopsis}, which backscatters Bluetooth to create Wi-Fi signals. Our results show that backscattering 2~Mbps Wi-Fi signals using Bluetooth consumes only 28~$\mu$W. Finally, to demonstrate the generality of our approach, we also show the feasibility of generating ZigBee signals by backscattering Bluetooth transmissions.

To show the potential of our design, we implement proof-of-concepts for the three applications shown in Fig.~\ref{fig:applications}. We build a contact lens form-factor antenna and evaluate it in-vitro to demonstrate that it can communicate with commodity devices. We also build an implantable neural recording interface antenna and evaluate it in-vitro using muscle tissue. \textred{Finally, to demonstrate the applicability of our techniques beyond medical implants, we create credit card form-factor devices that can communicate with each other using Bluetooth transmissions as the RF signal source for backscatter.}



\section{System Design}
We use backscatter to transform transmissions from Bluetooth devices into Wi-Fi signals. In this section, we first provide an overview of Bluetooth and Wi-Fi physical layers and then describe how to create single-tone transmissions using Bluetooth devices. We then show how to create an 802.11b signal from this single tone Bluetooth transmission. Finally, we outline our design for bi-directional communication using OFDM Wi-Fi as an AM modulator.

\begin{figure}[!t]
        \includegraphics[width=\columnwidth]{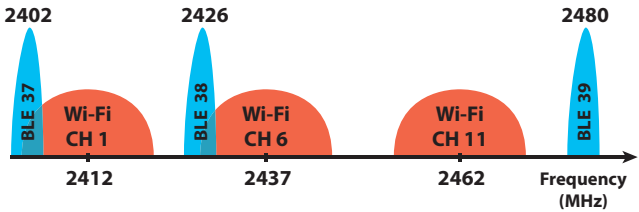}
\vskip -0.1in
  \caption{{\bf Wi-Fi versus Bluetooth.} The blue frequencies represent the three Bluetooth advertising channels and the red frequencies represent the three Wi-Fi channels.}
\label{fig:versus}
\vskip -0.15in
\end{figure}

\subsection{Bluetooth Versus Wi-Fi}
\label{sec:versus}

\vskip 0.05in\noindent{\bf Bluetooth.} Bluetooth low energy (BLE) devices use the advertisement channels to broadcast information about their presence and to initiate connections. Once the connection is established with a nearby Bluetooth device, they communicate by hopping across the 36 data channels spread across the 2.4~GHz ISM band. The three advertisement channels labeled as channels 37, 38 and 39 are shown in Fig.~\ref{fig:versus}. Since transmissions on data channels require establishing a connection with another device, we focus on Bluetooth advertisement channels where we can broadcast packets. Bluetooth LE uses Gaussian Frequency Shift Keying (GFSK) modulation with a bandwidth of 2~MHz.  Specifically, a `1' (`0') bit is represented by a positive (negative) frequency shift of approximately 250~kHz from the center frequency. The resulting FSK signal is then passed through a Gaussian filter to achieve the desired spectral shape.

\vskip 0.05in\noindent{\bf Wi-Fi.} While Wi-Fi supports a suite of standards, we focus on 802.11b for the purpose of backscatter. Wi-Fi operates on three non-overlapping channels, each 22~MHz wide. To create 1 and 2~Mbps transmissions, 802.11b first XORs each data bit with a Barker sequence to create a sequence of eleven coded bits for each incoming data bit, which it then modulates using DBPSK and DQPSK. To create 5.5 and 11~Mbps transmissions, 802.11b uses CCK where each block of four incoming bits is mapped to 8-bit code words, which are then transmitted using DBPSK and DQPSK.

\subsection{Bluetooth as an RF source}
\label{sec:bluetooth}
The high level idea is to transform a Bluetooth chip into a single tone transmitter, i.e., with constant amplitude and frequency.  We leverage two insights about GFSK modulation used in Bluetooth. First, Bluetooth uses two frequencies to encode the zero and one data bits. Thus, if we could transmit a stream of constant ones or zeros, we create a single frequency tone. Second, passing a single tone through the Gaussian filter used by Bluetooth does not change its spectral properties since the filter only smooths out abrupt changes to the frequency. Thus, if we could get the Bluetooth chipsets to transmit a continuous stream of zeros or ones, then we effectively create a single tone. Achieving this in practice requires us to address two key challenges: data whitening and the BLE packet structure.

\begin{figure}[!t]
        \includegraphics[width=0.9\columnwidth]{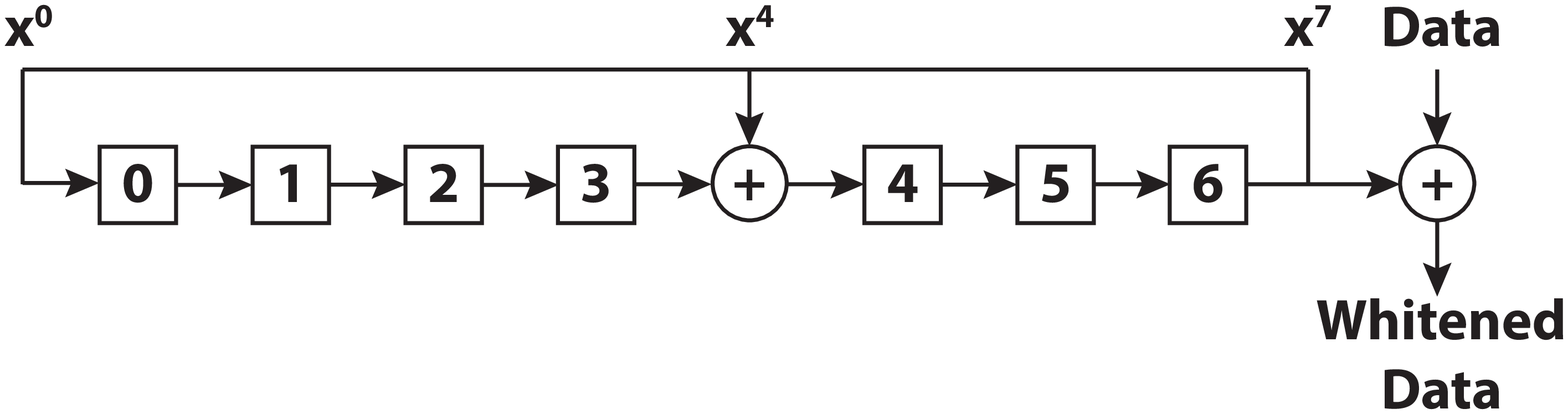}
\vskip -0.15in
  \caption{Shift registers used in Bluetooth to perform data whitening.}
\vskip -0.1in
\label{fig:ble_scrambler}
\end{figure}

\vskip 0.05in\noindent{\bf Data whitening.} While our goal is to create long sequences of either zeros or ones, Bluetooth uses data whitening to avoid such sequences so as to enable accurate timing recovery on a Bluetooth receiver. Specifically, Bluetooth uses the 7-bit linear feedback shift register circuit in Fig.~\ref{fig:ble_scrambler} with the polynomial $x^7 + x^4 +1$. Given an initial state, the circuit outputs a sequence of bits that are used to whiten the incoming data by XORing the data bits with the bits output by the circuit. We reverse this whitening process to create the desired sequence of ones or zeros. To do this, we observe that given an initial state for each of the registers, we can deterministically generate the whitening sequence.  We also note from the Bluetooth specification that the shift registers are initialized using the Bluetooth channel number. In particular, Bluetooth initializes the zeroth register to a one and the rest of the six registers to the binary representation of the Bluetooth channel number. For instance, while transmitting on the Bluetooth advertising channel 37, the zeroth register in Fig.~\ref{fig:ble_scrambler} is set to 1 and the rest are set to the binary representation of 37. Thus, given an advertising channel, we can initialize the Bluetooth whitening algorithm and compute the whitening sequence. The data bits are then set to the same bits in the whitening sequence or their bit complement to generate long sequences of zeros or ones respectively. In~\xref{sec:blue-results}, we show this process works with unmodified Bluetooth chipsets.

\begin{figure}[!t]
        \includegraphics[width=0.9\columnwidth]{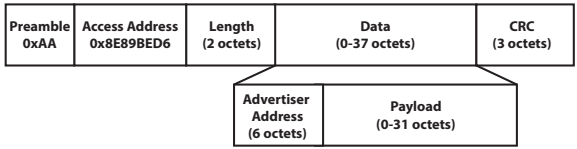}
\vskip -0.05in
  \caption{Structure of a Bluetooth advertising packet.}
\vskip -0.15in
\label{fig:ble_packet}
\end{figure}

\vskip 0.05in\noindent{\bf Link-layer packet structure.} The above discussion assumes that we can control all the bits in a Bluetooth packet; however an advertising packet has fields including the preamble and access address, shown in Fig.~\ref{fig:ble_packet}, that cannot be arbitrarily modified. The preamble is fixed to an alternating sequence of zeros and ones, and the access address is set to {\it 0x8E89BED6} for advertising packets. This is followed by a length field and an advertiser address field. Finally, the packet has the data payload and a 3-byte CRC. Of the above fields, only the data payload can be set to arbitrary values.\footnote{The Android API only allows 24 of the 32 bytes in the payload to be arbitrarily set.} Therefore, our system takes the following steps: 1) we use the Bluetooth preamble, access address and the header (56~$\mu$s in total) to enable Bluetooth packet detection at the backscattering device using an ultra-low power envelope detection circuit, 2) we estimate the beginning of the payload and start backscattering using the techniques introduced in~\xref{sec:genwifi} to generate Wi-Fi packets, and 3) we complete the Wi-Fi transmission before the start of the Bluetooth CRC field. Since the Bluetooth Bluetooth CRC is being transmitted on a different channel than the generated Wi-Fi packet, it does not affect the Wi-Fi receiver.

The above design requires estimating the beginning of the Bluetooth packet.  Our receiver uses an envelope detector, similar to prior ultra-low power designs~\cite{allsee-nsdi14,abc-sigcomm2013}, for energy detectors. Our energy detection circuit can be customized to only trigger for Bluetooth transmitter up to 8--10 feet, to prevent false positives. Energy detection however does not allow us to accurately detect the beginning of the Bluetooth packet since we do not decode the Bluetooth preamble and hence cannot use typical synchronization techniques. This results in inaccurate estimates of the beginning of the payload. Our implementation uses a guard interval of 4~$\mu$s in our estimate of the start of the payload to address this.


\subsection{Generating Wi-Fi  using backscatter}
\label{sec:genwifi}
We describe how to generate 802.11b signals by backscatter the single-tone from the Bluetooth device. We first present our design that can create a frequency shift to the incoming single tone signal. We then show how to synthesize 802.11b signals from this shifted tone.

\subsubsection{Single Sideband Backscatter Design}
\label{sec:singleside}
While existing sideband modulation techniques have been recently used to create frequency shifts~\cite{bluetoothbackscatter, nsdi16}, they also create a mirror copy. These works demonstrate that modulating the radar cross-section of an antenna effectively multiples a single tone signal by the modulating signal. Since, $2\sin(f_c t)\sin(\Delta ft) = \cos((f_c-\Delta f)t) - \cos((f_c+\Delta f)t)$, modulating the antenna at $\Delta f$ in the presence of the incoming single-tone carrier signal, $\sin(f_c t)$, creates the desired shift $\cos((f_c+\Delta f)t)$. However, it also creates the mirror copy $\cos((f_c-\Delta f)t)$. This is problematic when used with Bluetooth: both the advertising channels 37 and 39 are at either end of the ISM band, as shown in Fig.~\ref{fig:versus}. Thus, creating any frequency shifts to the corresponding Bluetooth signal will create a mirror copy outside the ISM band. The second advertising channel, 38, overlaps with Wi-Fi channel 6 and is close to Wi-Fi channel 1, and hence can create strong interference to the weak backscattered Wi-Fi signals. Further, generating packets on Wi-Fi channel 11 using advertising channel 39 shown in Fig.~\ref{fig:versus}, would again create a mirror copy that lies outside the ISM band. Thus, existing sideband modulation techniques cannot be used by \name\ on any of the BLE/Wi-Fi channels.

\vskip 0.05in\noindent{\bf Our Solution.} We present the first single sideband backscatter architecture that produces a frequency shift on only one side of the single tone Bluetooth transmission. Our intuition is to emulate radios. Specifically, radios use 2.4~GHz oscillators to generate the orthogonal signals, $\cos(2\pi f_c t)$ and $\sin(2\pi f_c t)$. These are multiplied with digital in-phase, $I(t)$, and quadrature phase components, $Q(t)$, to create the signal: $$[I(t)+j Q(t)]\times[\cos(2\pi f_c t) + j\sin(2\pi f_c t)]$$ By setting $I(t)$ and $Q(t)$ to $\cos(2\pi \Delta ft)$ and $\sin(2\pi \Delta ft)$, radios can easily create the desired shifted signal, $e^{j2\pi (f_c +\Delta f)t}$, without any mirror copies. The challenge is that we cannot use oscillators running at 2.4~GHz since they consume significant amounts of power~\cite{razavi1998rf}.

Our insight is that mathematically we can imitate the above operations using complex impedances on the backscatter device without 2.4~GHz oscillators. Say, we could create the complex signal $e^{j2\pi\Delta ft}$, then backscattering such a complex signal with the incoming single-tone Bluetooth transmission, $\cos(2\pi f_c t)$, results in:

\begin{eqnarray*}
  e^{j2\pi\Delta ft} \cos(2\pi f_c t) = \frac{1}{2} (e^{j2\pi(f_c+\Delta f)t} + e^{j2\pi(-f_c+\Delta f)t})
\end{eqnarray*}
The first term is the desired shifted signal while the second term has a negative frequency and does not occur in practice. Thus, the above operation creates the desired shift without a mirror copy. So if we can create the complex signal $e^{j2\pi\Delta ft}$ using backscatter, we can achieve single-sideband backscatter modulation. The desired signal can be written as,

\begin{eqnarray}
\label{eqn:1}
e^{j2\pi\Delta  ft} = \cos(2\pi\Delta ft) + j\sin(2\pi\Delta ft)
\end{eqnarray}
To create this on a backscatter device, we generate the sin/cos terms using square waves. We then use complex impedances at the switch to generate the desired complex values.

\vskip 0.05in\noindent{\it Step 1.}  We approximate the sin/cos terms in Eq.~\ref{eqn:1} using square waves alternating between +1 to -1,\footnote{ Since digital operations are on 0 and 1 bits, in practice we perform step 1 using a square wave between 1 and 0 instead of +1 and -1. This is however a straightforward mapping with a DC offset. For simplicity however, we explain our design using +1 and -1.} at a frequency of $\Delta f$. From Fourier analysis a square wave at $\Delta f$ can be written as $\frac{4}{\pi}\sum_{n=1,3,5,\cdots}\frac{1}{n} \sin(2\pi n\Delta ft)$. The first harmonic is the desired sine term while the third and fifth harmonic have a power of $\frac{1}{n^2}$ which are 9.5~dB and 14~dB lower than the first. Since all 802.11b bit rates can operate at SNRs lower than 14~dB, such an approximation is sufficient for our purposes. To generate the cosine term, we time shift this square wave by a quarter of the time period. This square wave can easily be generated by clocking the switch and the digital operations at multiples of the desired offset, $\Delta f$.

\vskip 0.05in\noindent{\it Step 2.} Now that we have approximated the sin/cos terms to be either +1 or -1, Eq.~\ref{eqn:1}  can take one of four values: {\it 1+j, 1-j, -1+j,} and {\it -1-j}. We create these complex values by changing the impedance of the backscatter hardware. RF signals are reflected when they cross two materials that have different impedances. Since the impedance of an antenna is different from the medium around it, a fraction of the incident RF signals get reflected off the antenna. Backscatter works by creating an additional impedance boundary between the antenna and the backscatter circuit. Specifically, given an incoming signal $S_{in}$, the reflected signal from the backscatter device is given by,

\begin{equation*}
S_{out} = \frac{Z_a - Z_c}{Z_a  +Z_c} S_{in}
\end{equation*}
Here $Z_a$ and $Z_c$ are the impedance of the antenna and the backscatter circuit respectively. In traditional backscatter, the impedance of the backscatter circuit is set to either $Z_a$ or $0$ corresponding to no reflection or maximum reflection of the incoming signal. We note however that the impedance of the backscatter circuit can be set to complex values by changing the inductance of the circuit~\cite{fullduplex-mobicom2014,matt-qambacksactter}. Specifically, at the frequency $f$, the impedance of the backscatter circuit can be written as $j2\pi fL$ where the inductance is $L$. Thus, by changing this inductance, we can create complex values for the fraction, $\frac{Z_a-Z_c}{Z_a+Z_c}$. Specifically, to get the four desired complex values, {\it 1+j, 1-j, -1+j, -1-j}, we set the impedances of the backscatter circuit to $\frac{-j}{2+j}Z_a$, $\frac{j}{2-j}Z_a$, $\frac{2-j}{j}Z_a$ and $\frac{2+j}{-j}Z_a$ respectively.  The antenna impedance, $Z_a$, is typically tuned to 50 ohms. By switching between these impedance states, we can generate the desired complex signal $e^{j2\pi \Delta ft}$ and hence achieve single-sideband backscatter modulation. \textred{Optimizing for these constraints, for our 2.4~GHz backscatter devices we used a 3~pF capacitor, open impedance, 1~pF and 2~nH to achieve the four complex values.}


\subsubsection{Synthesizing 802.11b signals}
\label{sec:ssb_iq}
A Wi-Fi signal can be written as, $(I_{wifi}(t) + jQ_{wifi}(t))e^{j2\pi f_{c}t}$, where $I_{wifi}(t)$ and $Q_{wifi}(t)$ correspond to the in-phase and quadrature-phase components for the baseband Wi-Fi signal. This can be rewritten as,
\begin{eqnarray*}
(I_{wifi}(t) + jQ_{wifi}(t)) e^{j2\pi\Delta ft} e^{j2\pi f_{bluetooth}t}
\end{eqnarray*}

Thus, to generate Wi-Fi signals, we need to create $(I_{wifi}(t) + jQ_{wifi}(t))e^{j2\pi\Delta ft}$ using backscatter. We have already demonstrated earlier how to generate $e^{j2\pi\Delta ft}$, so we can multiply it with the in-phase and quadrature-phase components of 802.11b to generate Wi-Fi signals. As described in~\xref{sec:versus}, 802.11b signals use DSSS/CCK coding that creates coded bits that are then modulated using either DBPSK or DQPSK. Thus, if we can show that we can transmit DBPSK and DQOPK, we can create 802.11b signals.


\vskip 0.05in\noindent{\bf DBPSK.} In this case, the one and zero bits are represented as +1 and -1, which translates to always setting $Q_{wifi}(t)$ to zero and $I_{wifi}(t)$ to either +1 or -1. Since $e^{j2\pi\Delta ft}$ takes the values in the set {\it \{1+j, 1-j, -1+j, -1-j\}}, multiplying it with +1 or -1 results in values within the same set, which we know how to generate using our impedance values. Thus, we can generate \label{eqn:2}  DBPSK modulation and hence achieve 1 and 5.5~Mbps 802.11b transmissions.

\vskip 0.05in\noindent{\bf DQPSK.} In this case, both $I_{wifi}(t)$ and $Q_{wifi}(t)$ are set to either +1 or -1. Thus, the baseband Wi-Fi signal can take one of the following values: {\it \{1+j, 1-j, -1+j, -1-j\}}. Multiplying this with $e^{j2\pi\Delta ft}$ which takes one of the following values {\it \{1+j, 1-j, -1+j, -1-j\}}, results in one of these four normalized values: {\it \{1,-1,j,-j\}}. We observe that {\it\{1,-1,j,-j\}} and {\it \{1+j, 1-j, -1+j, -1-j\}} (which are the four impedance values generated in our backscatter hardware) are constellation points that are shifted by $\pi/4$. Since 802.11b uses differential QPSK, we ignore this constant phase shift of $\pi/4$ and instead map them to the four complex impedance values generated by our hardware. Wi-Fi receivers ignore this constant phase shift since the bits are encoded using differential phase modulation. This allows us to generate DQPSK modulation and the corresponding 2 and 11~Mbps 802.11b transmissions.

\begin{figure}[!t]
    \includegraphics[width=\columnwidth]{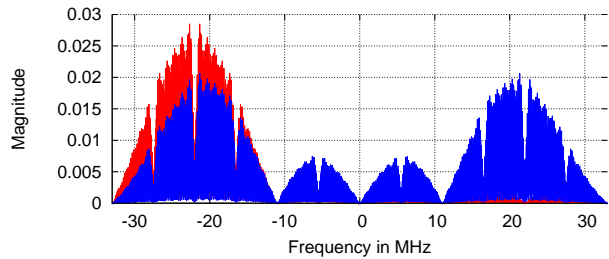}
\vskip -0.1in
  \caption{{ Comparing single sideband backscatter (red signal) to prior double sideband backscatter (blue signal).}}
\vskip -0.1in
\label{fig:ssb_spectra}
\end{figure}

Fig.~\ref{fig:ssb_spectra} shows the frequency spectrum for the backscatter generated WiFi signals at 2~Mbps using our single-sideband backscatter with an arbitrary frequency shift of 22~MHz. For comparison, we also plot the spectrum using prior sideband backscatter approaches~\cite{nsdi16}. The plots show that prior sideband backscatter designs create a strong mirror copy on the other side of the single tone. However, single-sideband backscatter, introduced in this paper, eliminates this mirror copy and hence improves spectral efficiency.

\subsubsection{Practical Design Considerations}

The Bluetooth advertising packet can have a payload up to 31 bytes or 248~$\mu$s. Since Wi-Fi packets at different bit rates occupy the channel for different times, this translates to different packet sizes. At 2, 5.5 and 11~Mbps the Wi-Fi payload can be 38, 104, and 209 bytes within a single Bluetooth advertising packet. Given its size, we however cannot fit a 1~Mbps Wi-Fi packet in a single Bluetooth advertising packets. \textred{We note that Bluetooth data transmissions, which are sent at faster rates and can last up to 2~$ms$, would enable 1~Mbps packets and greater overall throughput.} We focus on Bluetooth advertising packets in this paper since they are easier to control on commodity devices.

\vskip 0.05in\noindent{\it Additional Optimizations.}
Bluetooth does not perform carrier sense before transmitting. Further, the backscatter Wi-Fi packet is at a different frequency that could be occupied, resulting in a collision. Since Bluetooth advertisements are small and sent once every 20~ms, such collisions have a negligible impact on Wi-Fi which operates at a much finer time granularity. Collisions however are not desirable at the backscattering device; they would require the backscattering device to retransmit its data, consuming more energy.  We describe three optimizations that can reduce these collisions.

\vskip 0.05in\noindent 1) We could ensure that the Wi-Fi channel is unoccupied for the backscatter duration. Since most devices have both Wi-Fi and Bluetooth, they could coordinate and hence the Wi-Fi radio could schedule a $CTS\_to\_Self$ packet to be transmitted before the Bluetooth packet. $CTS\_to\_Self$ can reserve the channel for the duration of the Bluetooth packet, preventing other Wi-Fi devices from making concurrent transmissions. The ability to schedule $CTS\_to\_Self$ packets has been demonstrated in the past~\cite{tep,wifibackscatter-sigcomm2014} and requires driver and firmware access to the commodity device.

\vskip 0.05in\noindent 2) We leverage that the advertisement packets are sent on all Bluetooth advertising channels  one after the other, separated by a fixed duration $\Delta T$ (around 400~$\mu$s for TI Bluetooth chipsets).  Using this we imitate an RTS-CTS exchange: when Bluetooth transmits on channel 37, we backscatter an RTS packet on the desired Wi-Fi channel. If the channel is free, the Wi-Fi device responds with a CTS packet, effectively reserving Wi-Fi channel 11 for the next $2\Delta T + T_{Bluetooth}$, where $T_{Bluetooth}$ is the duration of the Bluetooth packet. The backscattering device detects the presence of this CTS using our peak detection hardware. It then transmits data packets on the desired Wi-Fi channel using the remaining advertising packets sent on channel 38 and 39 over the next $2\Delta T + T_{Bluetooth}$ seconds. 

\vskip 0.05in\noindent 3) To reduce the energy overhead of transmitting the RTS packet, we could instead transmit a data packet. If the Wi-Fi receiver can decode this packet, we would have exchanged useful data to the receiver. The Wi-Fi device can then send a $CTS\_to\_Self$ packet reserving the channel for the next $2\Delta T + T_{Bluetooth}$, which can then be used to backscatter additional Wi-Fi packets using the two remaining advertising packets, without collisions. This eliminates the energy overhead of sending a data-free RTS packet. Evaluating the effectiveness of this is not in the scope of this paper.

\vskip 0.05in\noindent\textred{Finally we note that having a large number of backscatter devices will not be affected by the RTS transmission since the multiple devices will be scheduled using the downlink transmissions before any of the backscatter devices even transmit.}

\subsection{Communication to Backscatter Device}
\label{sec:downlink}
Achieving communication to the backscattering device is a challenge because it cannot decode Wi-Fi and Bluetooth transmissions: Bluetooth uses frequency modulation while 802.11b uses phase modulation with DBPSK/DQPSK; so both have relatively constant amplitudes. Traditional receivers for such phase/frequency modulated signals require synthesis of a high frequency carrier that is orders of magnitude more power consuming than backscatter transmitter. In fact, existing ultra-low power backscatter receiver designs~\cite{abc-sigcomm2013,allsee-nsdi14,nsdi16} use amplitude modulation (AM) which does not require phase and frequency selectivity; unfortunately, Wi-Fi and Bluetooth radios do not support AM.

\begin{figure}[!t]
        \includegraphics[width=\columnwidth]{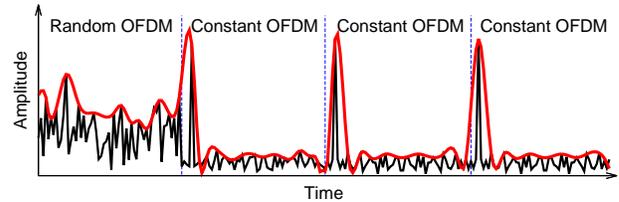}
\vskip -0.15in
\caption{{\bf Creating AM signals using OFDM.} We show four OFDM symbols (in black). Random OFDM symbols are constructed using IFFT over random modulated bits while constant OFDM symbols are constructed using the same modulated bits. The red line show the output of our peak detector receiver.}
\label{fig:ofdm_timedomain}
\vskip -0.15in
\end{figure}

\vskip 0.05in\noindent{\bf Our Approach.} We transform the payload of 802.11g packets into AM modulated signals. In 802.11g, each OFDM symbol is generated by taking an IFFT over QAM modulated bits to generate 64~time domain samples. Fig.~\ref{fig:ofdm_timedomain} shows a time-domain OFDM symbol created from random modulated bits, which we call {\it random OFDM}. We also show the symbol created when the IFFT is performed over constant modulated bits, which we call {\it constant OFDM}. The figure shows that while random OFDM symbols have the energy spread across the time samples, in constant OFDM the energy is in the first time sample and is zero elsewhere. This can be used to create an amplitude modulated signal. Constructing a constant OFDM symbol using Wi-Fi radios is however not straightforward due to scrambling, coding and interleaving. Next, we describe how to to achieve this goal.


\vskip 0.05in\noindent{\it Scrambler.} 802.11g data scrambling is performed by XORing the incoming data bits with the output of a 7-bit linear feedback shift register using the same polynomial shown in Fig.~\ref{fig:ble_scrambler}. Given the scrambling seed, the output sequence of the circuit is deterministic. According to the Wi-Fi standard, the scrambling seed is set to a pseudorandom non-zero value. In principle, this information should be available on the Wi-Fi hardware. Our experiments in~\xref{sec:res_downlink} reveal that a number of commercial Wi-Fi chipsets use a predictable sequence of scrambling seeds. Further, certain Atheros chipsets also allow us to set the scrambling seed to a fixed value in the driver. Using this, we reverse the effect of the scrambler.

\vskip 0.05in\noindent{\it Convolutional Encoder.} 802.11g uses convolutional encoding on the scrambled bits to be resilient to noise and interference. It uses a 1/2 rate convolutional encoder where two coded bits are output for each incoming scrambled bit. The higher 2/3 and 3/4 coding rates are obtained by dropping some of the 1/2 rate encoded bits. Specifically, given the scrambled bits, $b[k]$, the two encoded bits are,

\begin{eqnarray*}
C_1[k] = b[k] \oplus b[k-2] \oplus b[k-3] \oplus b[k-5] \oplus b[k-6]\\
C_2[k] = b[k] \oplus b[k-1] \oplus b[k-2] \oplus b[k-3] \oplus b[k-6]
\end{eqnarray*}
This is a 1-to-2 mapping which cannot generate every desired sequence of encoded bits. We observe however that if all the incoming scrambled bits are ones (zeros), then all the encoded bits are ones (zeros). Thus, we can generate a sequence of all zeros or all ones as encoded bits.

\vskip 0.05in\noindent{\it Interleaver.} The encoded bits are interleaved across different OFDM frequency bins to make adjacent encoded bits more robust to frequency selective channel fading. We note however that when using a sequence of all ones or zeros as our encoded bits, interleaving again results in a sequence of all ones or zeros and does not require us to perform any special operation.

\vskip 0.05in\noindent{\it Modulator.} The interleaved bits are modulated using BPSK, 4QAM, 16QAM or 64QAM. Since our interleaved bits are all ones or zeros within an OFDM symbol, the modulation operation results in using the same constellation point across all the OFDM bins, achieving our goal of creating an OFDM symbol constructed using a constant modulated symbol.

We note that OFDM symbols have pilot bits inserted in specific frequency bins, which cannot be controlled. This however does not significantly change the desired constant OFDM pattern since the fraction of pilot to data symbols is low. Also, 802.11g convolutional encoders have a delay length of $7$, i.e., the last six data bits from the previous OFDM symbol impact the first few encoded bits in the current OFDM symbol. This could be a problem when the constant OFDM symbol follows a random OFDM symbol. We address this by setting the last six data bits in the random OFDM symbol to ones and use 16/64 QAM to ensure that the random OFDM symbol will still result in a high amplitude signal.

\begin{figure}[!t]
        \includegraphics[width=\columnwidth]{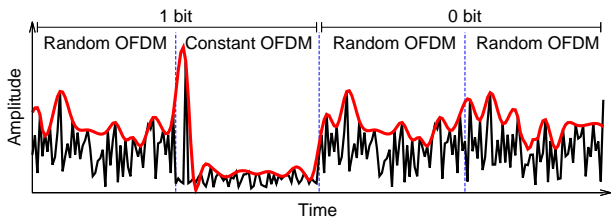}
\vskip -0.15in
\caption{{\bf Encoding bits with OFDM symbols.} The figure shows the encoding used to convey bits to the backscatter device.}
\label{fig:ofdm_timedomain1}
\vskip -0.15in
\end{figure}

\vskip 0.05in\noindent{\bf Encoding process.} Ideally, we can encode a 1 and 0 bit by random and constant OFDM symbols respectively.  Fig.~\ref{fig:ofdm_timedomain} shows why this is a problem: constant OFDM symbols have a peak at the beginning of the time-domain symbol. This is a problem since we use a passive peak detection receiver  that tracks the peaks in this signal (shown in red). This creates false peaks at the beginning of each constant OFDM symbol, which can confuse the receiver when there are consecutive constant OFDM symbols. To avoid this, we encode each bit with two OFDM symbols as shown in Fig.~\ref{fig:ofdm_timedomain1}. A one bit is represented by a random OFDM symbol followed by a constant OFDM symbol, while a zero bit is represented as two random OFDM symbols. Since each 802.11g OFDM symbol is 4~$\mu$s, this achieves a bit rate of $125$~kbps.

Finally, OFDM symbols have a cyclic prefix where the last few time samples are repeated at the beginning. Since the cyclic prefix in the case of a constant OFDM symbol is all zero, this could create a glitch. To avoid this, we pick the preceding random OFDM symbol such that its last time sample has a high amplitude. This ensures that the peak detector circuit sees a high peak at the end of the first OFDM symbol and does not create a glitch during the cyclic prefix.

\subsection{Putting it all together}
We use a query-reply protocol. The Wi-Fi device queries the backscatter device using the reverse channel in~\xref{sec:downlink}. The backscatter device responds using the backscatter channel in~\xref{sec:genwifi}. This design works with multiple backscatter devices since the Wi-Fi device can query them one after the other.

\section{FPGA and IC Design}
We began by developing the hardware on an FPGA platform to characterize the system and build proof of concept applications. We then translated the design into an IC and used it to quantify the power consumption.

\noindent\textbf{FPGA design.} Our system has two components: the RF front end and the baseband digital circuit. The front end consists of a backscatter modulator and a passive receiver. We isolate the receiver from the antenna using a single pole double throw (SPDT) switch that switches between transmit and receive modes. The backscatter modulator switches between four impedance states and is implemented using a cascaded two-stage SPDT switch network. We use HMC190BMS8 SPDT switches both for isolating the transmitter and receiver and in the backscatter modulator. The front end is implemented on a low loss Rodgers 4350 substrate~\cite{hmc190bms8}. As explained in~\xref{sec:genwifi}, the impedances connected to the four terminals of the switch network are a 3~pF capacitor, open impedance, a 1~pF capacitor and a 2~nH inductor which achieve the four desired complex values. The receiver is an energy detector consisting of passive analog components and a comparator to distinguish between the presence and absence of energy. We replicate the receiver described in ~\cite{allsee-nsdi14, abc-sigcomm2013, nsdi16}.

The 802.11b scrambling, DSSS/CCK encoding, CRC encoding, DQSPK encoding and single-side band backscatter were implemented in Verilog \textred{and translated onto the DE1 Cyclone II FPGA development board by Altera~\cite{altera_de1}}. We implement a 35.75~MHz shift which we found to be optimal for rejecting the interference from the Bluetooth RF source. The digital output of the FPGA was connected to the backscatter modulator and  the energy detector was fed to its digital input. A 2~dBi antenna was used on the \name\ device.

\begin{figure*}[!t]
        \subfigure[TI CC2650]
        {
        \includegraphics[width=.3\textwidth]{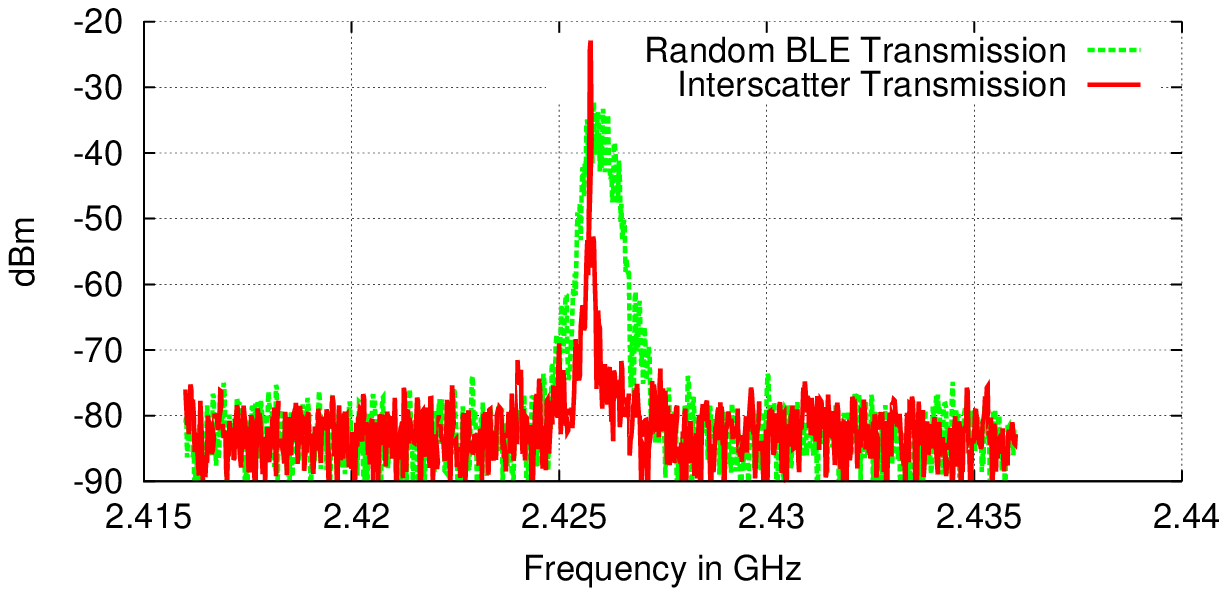}
        }
        \subfigure[Galaxy S5 Smartphone]
        {
        \includegraphics[width=.3\textwidth]{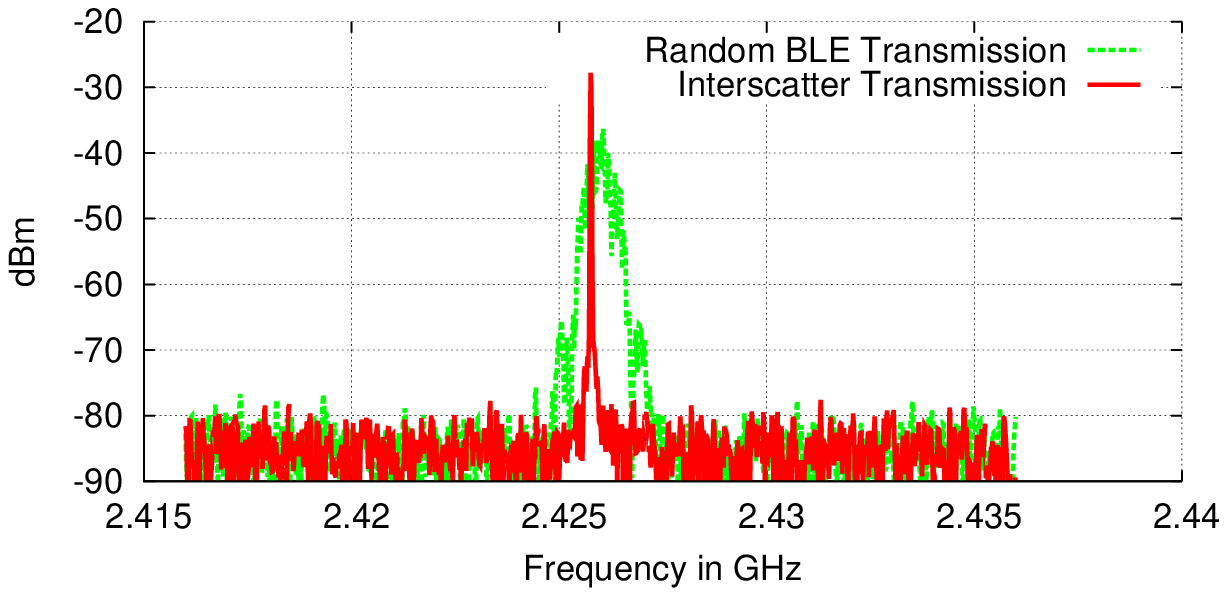}
        }
        \subfigure[Moto360 (2nd gen) Smartwatch]
        {
        \includegraphics[width=0.3\textwidth]{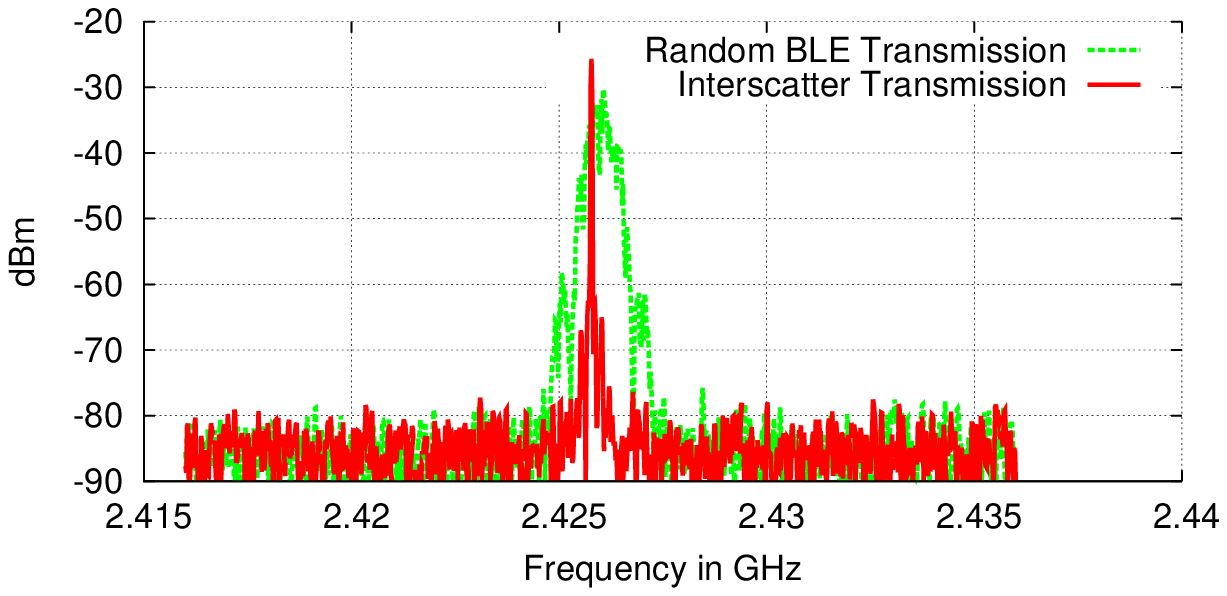}
        }
\vskip -0.1in
        \caption{{\bf Generating Bluetooth single tone.} The red lines show Bluetooth transmissions with random application data and the green lines show the single-tone transmission created by \name.}
\label{fig:spectra}
\vskip -0.15in
\end{figure*}

\noindent \textbf{IC design.}  As CMOS technology has scaled, the power consumption of digital computation has decreased significantly~\cite{koomey}. Unfortunately, active radios require power hungry \textit{analog} components which operate at RF frequencies and do not scale in either power or area.  Interscatter relies exclusively on baseband digital computation with no components operating at RF frequencies, so it can leverage CMOS scaling for ultra-low power operations. We implement \name\ on the TSMC 65~nm low power CMOS technology node. For context, Atheros AR6003~\cite{atheros_ar6003} chipsets released in 2009 used 65~nm CMOS and so, it is a fair comparison with industry standard. The \name\ IC implementation can be broken down into the following three main components described in detail below: the frequency synthesizer, baseband processor, and backscatter modulator.

\noindent \textit{Frequency synthesizer.} This block takes a frequency reference as an input and generates the 802.11b baseband 11~MHz clock as well as the four phases of the 35.75~MHz frequency clock required for backscatter. \textred{We use an integer N charge pump and ring oscillator based PLL to generate a 143~MHz clock which is fed to a Johnson counter to generate the four clocks for the 35.75~MHz frequency shift each 90\degree~out of phase.} The same 143~MHz clock is divided by 13 to generate the 11~MHz baseband clock. Thus, 11~MHz and 35.75~MHz are phase synchronized and we avoid glitches. This block consumes 9.69 $\mu W$ of power.

\noindent\textit{Baseband processor.} This block takes the payload as the input and generates the baseband 802.11b packet. We use the same Verilog code which was verified on the FPGA and synthesize the transistor level implementation using the Design Compiler tool by Synopsis~\cite{synopsis}. This block has a power consumption of 8.51 $\mu W$ for 2~Mbps Wi-Fi transmissions.

\noindent\textit{Backscatter modulator.} We implement our single sideband backscatter in the digital domain by independently generating the in-phase and quadrature phase components. We take the two bit output of the baseband processor and feed it to two multiplexers that map the four phases of the 35.75~MHz carrier to corresponding in-phase and quadrature-phase components. Then, at each time instant, the in-phase and quadrature phase components are mapped to the four required impedance states in~\xref{sec:ssb_iq}. We use CMOS switches to choose between open, short, capacitive and inductive states. The single side band backscatter modulator consumes 9.79 $\mu W$. {\it In total, generating 2~Mbps 802.11b packets consumes 28 $\mu W$.}

\section{Evaluation}
We first check if we can create single tone transmissions on various Bluetooth platforms. We then measure the communication range and packet loss rate for the Wi-Fi packets generated by backscattering Bluetooth. Next, we evaluate the efficacy of our single sideband backscatter architecture. We then evaluate the communication link from an 802.11g transmitter to our low-power receiver. Finally, we demonstrate the feasibility of generating ZigBee transmissions.

\subsection{Generating Bluetooth single tone}
\label{sec:blue-results}
We run experiments with three different Bluetooth devices: the Texas Instruments CC2650, a Moto 360 2nd gen smart watch, and a Samsung Galaxy S5. We connected the exposed antenna connector on the TI chipsets to a spectrum analyzer and recorded data during the payload section of a Bluetooth packet. The Android platforms do not expose antenna connectors, so instead we performed the same experiment by recording the signal through a a 2~dBi monopole antenna connected to the spectral analyzer. We set the application layer data as described in~\xref{sec:bluetooth} to create a single tone. Fig.~\ref{fig:spectra} shows the spectrum of random Bluetooth transmissions in green. It also shows the same spectrum (in red) after performing the operations described in~\xref{sec:bluetooth} for comparison. The plots show that we can create single tone transmissions from commodity Bluetooth devices.

  \subsection{Measuring the Wi-Fi RSSI}
  We measure Wi-Fi RSSI for different distances between the backscatter device, Bluetooth transmitter and Wi-Fi receiver. We first fix the distance between the backscatter device and the Bluetooth transmitter. We then move the Wi-Fi receiver perpendicular from the mid-point of the line connecting the other two devices and measure the RSSI of the backscattered Wi-Fi packets reported by the Wi-Fi receiver. We set the backscatter device to generate 2~Mbps Wi-Fi packets on channel 11. The Bluetooth transmitter sends advertising packets with a 31 byte payload on BLE channel 38, once every 40~ms. We use a TI Bluetooth device and an Intel Link 5300 Wi-Fi card as our Bluetooth transmitter and Wi-Fi receiver respectively. We use four power values at the Bluetooth transmitter: 1) 0~dBm which is the typical configuration for Bluetooth devices, 2) 4~dBm, which is supported on Samsung S6~\cite{fcc-samsungs6} and One Plus 2 ~\cite{fcc-oneplus2}, 3) 10~dBm, which is supported by Samsung Note 5~\cite{fcc-samsungnote5} and iPhone 6~\cite{fcc-iphone6}, and 4) 20~dBm which is supported by class 1 Bluetooth devices.

\begin{figure}[!t]
        \subfigure[1~feet between the backscatter device and Bluetooth transmitter]
        {
        \includegraphics[width=\columnwidth]{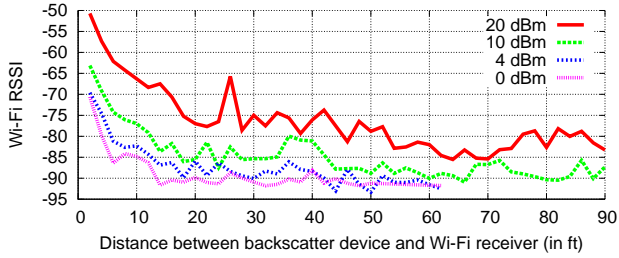}
        }
        \subfigure[3 feet between the backscatter device and Bluetooth transmitter]
        {
        \includegraphics[width=\columnwidth]{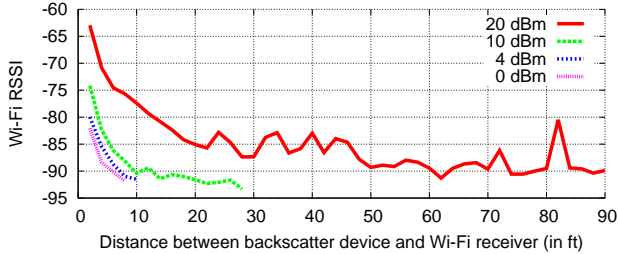}
        }
  \caption{{\bf Measuring the Wi-Fi RSSI.} RSSI versus distances between the backscattering device, Bluetooth transmitter and Wi-Fi receiver at different transmit power values.}
\label{fig:rssi_dist}
\vskip -0.15in
\end{figure}

\begin{figure}[!t]
    \includegraphics[width=\columnwidth]{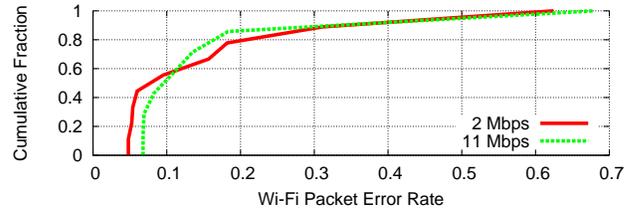}
\vskip -0.1in
  \caption{{\bf Wi-Fi packet error rate.} We measured packet error rate for backscatter-generated Wi-Fi packets at both 2 and 11 Mbps.}
\vskip -0.1in
\label{fig:per}
\end{figure}

Fig.~\ref{fig:rssi_dist} shows the results when the Bluetooth transmitter and the backscatter device are 1 and 3 feet apart. We pick these distances since we expect the user to use their mobile devices close to the backscatter device they intend to interrogate.  The x-axis plots the distance between the Wi-Fi receiver and the backscatter device while the y-axis plots the reported Wi-Fi RSSI values. The figure shows that:

\begin{Itemize}
\item At higher transmit powers, the range at which Wi-Fi packets are reported is high --- with 20~dBm, we achieve  a range of around 90~feet.  This is because higher transmit powers translate to high-powered Bluetooth signals at the backscatter device. This in turns increases the RSSI of the backscatter generated Wi-Fi packets.
\item For a similar reason, at larger distances between the Bluetooth transmitter and the backscatter device, the Wi-Fi RSSI and range are lower. We however note that the achieved ranges are more than sufficient for the target personal area networking applications shown in Fig.~\ref{fig:applications}.

\item While the RSSI values are lower than typical 802.11n/g deployment, the reported RSSI values are mostly sufficient for decoding 2~Mbps 802.11b transmissions given that theoretically 2~Mbps Wi-Fi transmissions require an SNR of only 6~dB for reliable communication.

\end{Itemize}
Finally, we compute the packet error rate (PER) observed on the backscatter generated Wi-Fi packets across the whole spectrum of RSSI values observed in our experiments. To do this, we configure the backscatter device to consecutively transmit 200 unique sequence numbers in a loop that we use to compute the error rate the Wi-Fi receiver. We compute the packet error rate for both 2 and 11~Mbps Wi-Fi transmissions. For 2~Mbps and 11~Mbps, we generate packets with a payload of 31 and 77~bytes respectively so as to fit within a Bluetooth advertising packet. Fig.~\ref{fig:per} shows a CDF of the observed packet loss rate. The figure shows that,
\begin{Itemize}
\item Both 2~Mbps and 11~Mbps transmissions have similar packet loss rates. This is because the Wi-Fi packet payload sizes that can fit within a Bluetooth advertising packet are small. Further, both our 2~Mbps and 11~Mbps Wi-Fi packets use preamble and header encoded at the same bit rate. Thus, we see similar packet error rates between the two transmissions. This provides an interesting design choice in our system, where given the small packet sizes, if we were to optimize for the number of retransmissions, it is better to use the higher Wi-Fi bit rates and transmit more bits. Exploring this and how it interacts with power consumption would be an interesting future direction.
\item The packet error rate was greater than 30\% for low RSSI values. In principle, we can leverage prior work in our community~\cite{marnello,mrd} that use diversity and coding to account for such packet losses in Wi-Fi networks.
\end{Itemize}

     \subsection{Efficacy of single sideband backscatter}
         We compare the single sideband backscatter design introduced in this paper with prior double sideband backscatter hardware designs~\cite{nsdi16}. We use a USRP to transmit a single tone carrier such that the unintended mirror copy generated by double sideband backscatter appears in Wi-Fi channel 6. We then configure a standard Wi-Fi transmitter-receiver pair on channel 6 and evaluate the effect of the backscatter interference by running iperf using TCP with the default Wi-Fi rate adaptation algorithm. We use a Linksys WRT54G Wi-Fi AP as the Wi-Fi transmitter and a Nexus 4 smartphone as the Wi-Fi receiver separated by 10~feet. We set the backscatter device to generate 2~Mbps Wi-Fi packets with 32 byte payloads, at a distance of 2~feet from the Wi-Fi receiver.

\begin{figure}[!t]
    \includegraphics[width=\columnwidth]{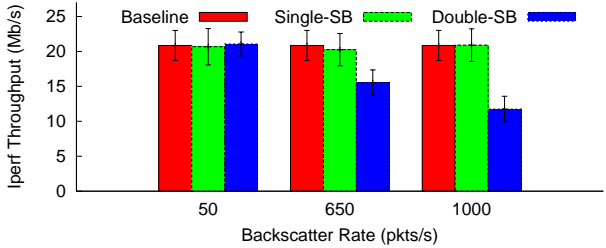}
  \caption{{\bf Efficacy of single sideband backscatter.} We compare our design with double sideband backscatter designs on the throughput of an iperf flow on a concurrent Wi-Fi transmitter-receiver pair. Baseline is the throughput in the absence of any backscatter device.}

\label{fig:interference}
\end{figure}

Fig.~\ref{fig:interference} shows the iperf throughput in the presence of single and double-sideband backscatter hardware. The x-axis shows the rate at which Wi-Fi packets are generated using backscatter. For comparison, we show the baseline throughput in the absence of both backscatter devices. The plot shows that,
\begin{Itemize}
\item When the backscattering device generates a small number of packets (50 pkts/s), it has negligible impact on the concurrent iperf flow. This is true with both single-sideband and prior double sideband backscatter designs. This is expected because the backscattered packets are small and are transmitted at a very low rate that they do not affect the throughput of concurrent Wi-Fi connections.
\item When we backscatter Wi-Fi packets at a higher rate, the iperf throughput is reduced with prior double sideband backscatter hardware. This is because it creates a mirror copy on Wi-Fi channel 6 that reduces the throughput of any other Wi-Fi connection. However, we see negligible impact on the iperf throughput with our single-sideband hardware. This demonstrates that our single-sideband backscatter design can double spectral efficiency.
\end{Itemize}

  \subsection{Communication in reverse direction}
  \label{sec:res_downlink}

\begin{figure}[!t]
    \includegraphics[width=\columnwidth]{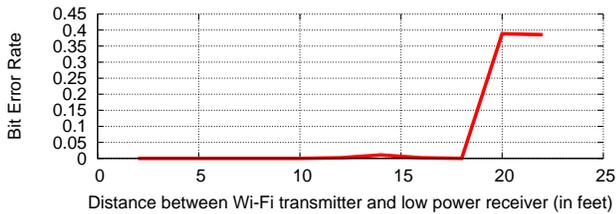}
\vskip -0.1in
  \caption{BER from  802.11g device to our low-power receiver.}
\vskip -0.15in
\label{fig:downlink}
\end{figure}

As described in~\xref{sec:downlink}, we create an AM modulated signal using OFDM by setting the appropriate modulated bits on each OFDM symbol. This however requires us to know the scrambling seed that is used by the Wi-Fi transmitter. We run experiments to track how different chipsets change the scrambling seed between packets by transmitting 802.11g packets at a bit rate of 36~Mbps from each Wi-Fi device. Since existing Wi-Fi receivers do not expose the scrambling seed, we instead use the gr-ieee802-11 package~\cite{usrp-802.11g} for GNURadio which implements the complete 802.11g receive chain to view the scrambling seed for each packet. Our experiments reveal that the AR5001G, AR5007G and AR9580 Atheros chipsets simply increment the scrambling seed by one between frames which is consistent with other studies~\cite{scrambler-seed}. We were also able keep the seed value fixed on ath5k cards by setting the GEN\_SCRAMBLER field in the AR5K\_PHY\_CTL register of the driver.

Next, we evaluate how our reverse link works with our low-power peak detector receiver. We transmit 36~Mbps 802.11g packets scrambled with a known seed. The Wi-Fi transmitter is configured to send a pre-defined repeating sequence of bits using the encoding in~\xref{sec:downlink}. We move our low-power receiver away from the Wi-Fi transmitter and compute the observed bit error rate at each location. Fig.~\ref{fig:downlink} shows the BER results as a function of the distance between the Wi-Fi transmitter and our peak detector receiver. The plot shows that despite the significant variability of OFDM, our peak detector achieves bit error rates less than 0.01 at distance of up to 18. Our current receiver  uses off-the-shelf components and has a sensitivity of -32~dBm for 160~kbps. This however can be improved in a custom IC implementation that could give us higher ranges.

\subsection{Generating ZigBee Using Backscatter}
  Finally, we show the feasibility of generating ZigBee signals by backscattering Bluetooth transmissions. ZigBee operates in the 2.4~GHz band over 16 channels where each channel is 5~MHz wide. At the physical layer, ZigBee achieves bit rates of 250~kbps using DSSS coding and offset-QPSK and has a better noise sensitivity than Wi-Fi. Since we have demonstrated that we can generate 802.11b that uses DSSS and QPSK, we adapt the techniques in~\xref{sec:genwifi} to also generate ZigBee-compliant packets.

\begin{figure}[!t]
    \includegraphics[width=\columnwidth]{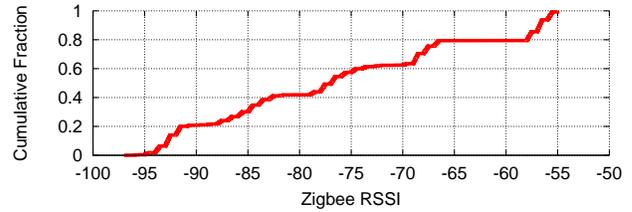}
\vskip -0.1in
  \caption{{\bf CDF of ZigBee RSSI.} We measure the RSSI of backscatter-generated ZigBee packets at five different locations.}
\vskip -0.15in
\label{fig:zigbee}
\end{figure}

To evaluate this, we use the TI CC2650 Bluetooth device as our Bluetooth transmitter on advertising channel 38 and set our backscatter device to generate packets on channel 14, i.e., 2.420~GHz. We use the TI CC2531 as our commodity ZigBee receiver to receive the packets generated by the backscatter device. We place the backscattering device two feet away from the Bluetooth transmitter and the ZigBee receiver at five different locations up to 15~feet from the backscatter device. Fig.~\ref{fig:zigbee} shows the CDF of the ZigBee RSSI values for various distance locations,  showing the feasibility of generating ZigBee packets with backscatter. We note that existing ZigBee transmitters consume tens of milliwatts of power when actively transmitting. In contrast, our backscatter based approach would consume tens of microwatts  while transmitting a packet and could be beneficial for short range communication with nearby ZigBee devices.

\section{Proof-of-Concept Applications}

This paper demonstrates backscatter communication on commodity devices, and in this section we evaluate prototypes of three novel applications enabled by these techniques. We evaluate only the communication aspect of our prototypes and consider fully exploring their security and usability issues outside the of scope of this work.

\subsection{Smart Contact Lens}
Smart contact lens systems~\cite{liao20123,yao2012contact} measure biomarkers like glucose, cholesterol and sodium in tears that can help with unobtrusive tracking for diabetic patients. The lens consists of a miniature glucose sensor IC, and an antenna. Although the power required for glucose sensing is minimal, real-time communication is power consuming and rules out conventional radios. Today these systems are limited to periodically sampling and storing the glucose data that is then transmitted sporadically using backscatter whenever the lens is next to a dedicated, powered, RFID-like reader.

\begin{figure}[!t]
        \subfigure
        {
          \includegraphics[width=.25\columnwidth]{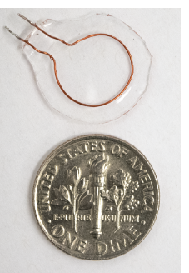}
        }
        \subfigure
        {
          \includegraphics[width=0.7\columnwidth]{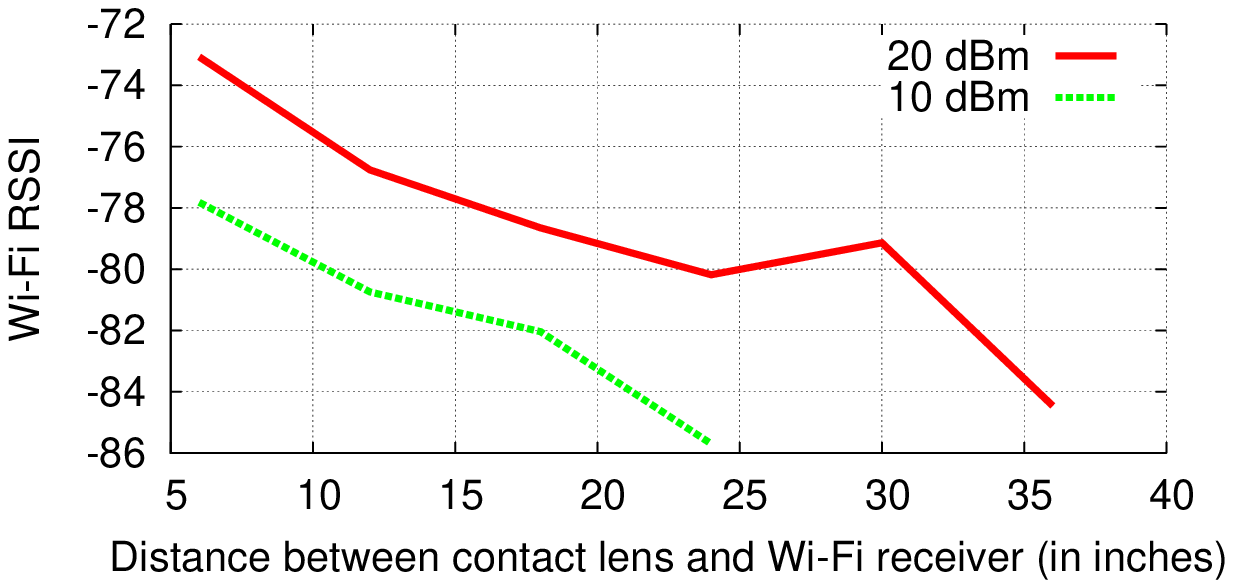}
        }
\vskip -0.1in
  \caption{{\bf RSSI with smart contact lens antenna prototype.}}
\vskip -0.15in
\label{fig:rssi_contact_lens}
\end{figure}

We leverage our design to show that a smart contact lens can communicate with commodity Wi-Fi and Bluetooth radios, without the need for dedicated reader hardware. We develop the form factor antenna prototype lens shown in Fig.~\ref{fig:rssi_contact_lens}. The prototype consists of a 1~cm diameter loop antenna similar to prior contact lens systems~\cite{liao20123, yao2012contact}. The antenna was built using 30~AWG wire and then encapsulated in a 200 $\mu$m thick layer of poly-dimethylsiloxane (PDMS) for biocompatibility and structural integrity. We then connect the antenna to our FPGA prototype. Unlike traditional antennas that have a 50 $\Omega$ impedance, small loop antennas have non-standard impedances and so we optimize the impedance of the backscatter switch network for the impedance of our loop antenna. We note that for ease of prototyping, we use FPGA hardware but an IC should conform to the contact lens size form factor, similar to prior backscattering lens designs~\cite{lens1}.

To evaluate the system, we immerse our antenna prototype in contact lens solution. We configure the TI Bluetooth chip to transmit every 40~ms and place it 12 inches from the lens. We use the Intel Wi-Fi card as a receiver and vary its distances to the lens antenna. The RSSIs follow a similar trend when using a Samsung Galaxy S4 as the receiver. We configure our FPGA to detect the Bluetooth beacons and backscatter 2~Mbps Wi-Fi packets. Fig.~\ref{fig:rssi_contact_lens} plots the Wi-Fi RSSI for different Bluetooth transmit power values. The plot shows that we can achieve ranges of more than 24~inches, demonstrating the feasibility of a smart contact lens that communicate directly with commodity radios. We note that the range is smaller than prior plots because the antenna is much smaller and is immersed in liquid leading to high signal attenuation. The range however is similar to prior smart contact lens with a dedicated RFID-like reader~\cite{lens1}.

\vfill
\subsection{Implanted Neural Recording Devices}
Implantable neural recording devices have recently demonstrated promising results towards use in brain-computer interfaces (BCIs)~\cite{leuthardt2004brain} that help paralyzed individuals operate motor prosthetic devices, command electronic devices, or even regain control of their limbs~\cite{hochberg2006neuronal}. These systems use either penetrating neural probes~\cite{bmi2} or a surface electrode grid~\cite{wyler1984subdural} that is implanted to collect local field potentials and electrocorticography (ECoG) signals~\cite{bmi2}. The recording sensors today consume around 2~$\mu$W/channel~\cite{wyler1984subdural}, with 8--64 parallel recording channels.

\begin{figure}[!t]
        \subfigure
        {
          \includegraphics[width=.25\columnwidth]{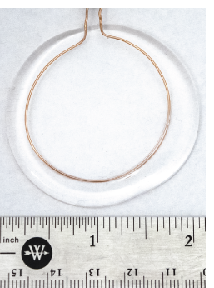}
        }
        \subfigure
        {
        \includegraphics[width=0.7\columnwidth]{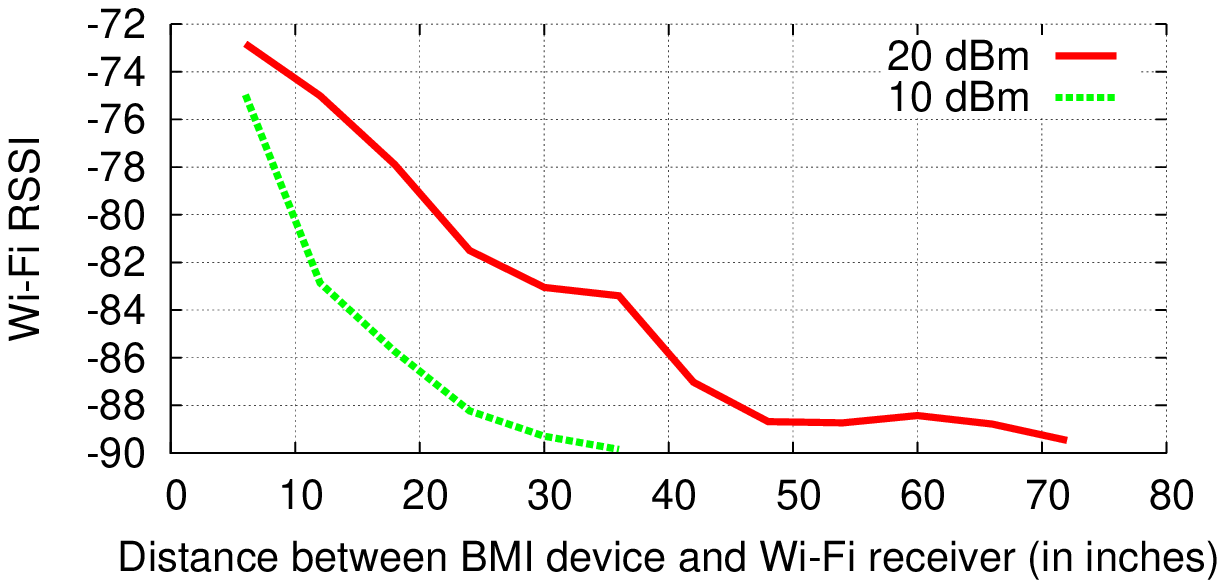}
        }
\vskip -0.1in
\caption{{\bf RSSI with implantable neural recording antenna.}}
\vskip -0.15in
\label{fig:rssi_bmi}
\end{figure}
Instead of using a custom backscatter reader as prior prototypes~\cite{bmi1,bmi2,bmi3}, we leverage our design to transmit the data directly to commodity radios. To demonstrate this, we built a form factor antenna for the implantable neural recording device shown in Fig.~\ref{fig:rssi_bmi}. We construct a 4~cm diameter full wavelength loop antenna using 16 AWG magnet wire and encapsulate it in a 2~mm thick layer of PDMS to isolate it from biological tissue.

These devices are typically implanted either on top of the dura beneath the skull or just under the skin inside the skull cavity. For our in-vitro evaluation, we choose muscle tissue whose electromagnetic properties are similar to grey matter at 2.4~GHz~\cite{gabriel1996dielectric}. We took a 0.75~inch thick pork chop and created a slit 0.0625~inch from the surface and inserted the antenna. We use the TI Bluetooth device as our RF source and place it at a distance of 3~inches from the meat surface. We use Intel Link 5300 Wi-Fi card on channel 11 as our Wi-Fi receiver; the RSSI results were similar with a Samsung S4. We set our FPGA prototype to synthesize 2~Mbps Wi-Fi packets. Fig.~\ref{fig:rssi_bmi} shows the RSSI for different Bluetooth powers. The plots show the feasibility of communicating with implanted devices, despite significant attenuation due to the muscle tissue. The range we achieve is better than the 1-2~cm target range for existing prototypes~\cite{bmi1, bmi2}. Finally, since Samsung Note 5 and iPhone 6 support 10~dBm Bluetooth transmissions, this demonstrates an implantable device that can communicate with commodity mobile devices.

\subsection{Card to Card Communication}
Finally, we can use the single-tone transmission from Bluetooth devices to enable communication between two passive cards similar to ambient backscatter~\cite{abc-sigcomm2013}, but without the need for strong TV transmissions. Since most users have Bluetooth-enabled devices (e.g., smartphone), this could be used for money transfer between credit cards, bus passes, splitting of checks, or transferring content to digital paper devices.

\begin{figure}[!t]
        \subfigure
        {
          \includegraphics[width=.25\columnwidth]{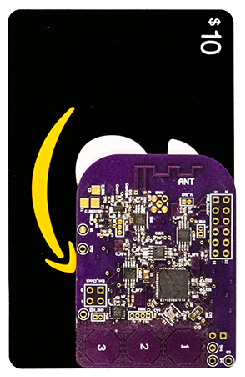}
        }
        \subfigure
        {
        \includegraphics[width=0.7\columnwidth]{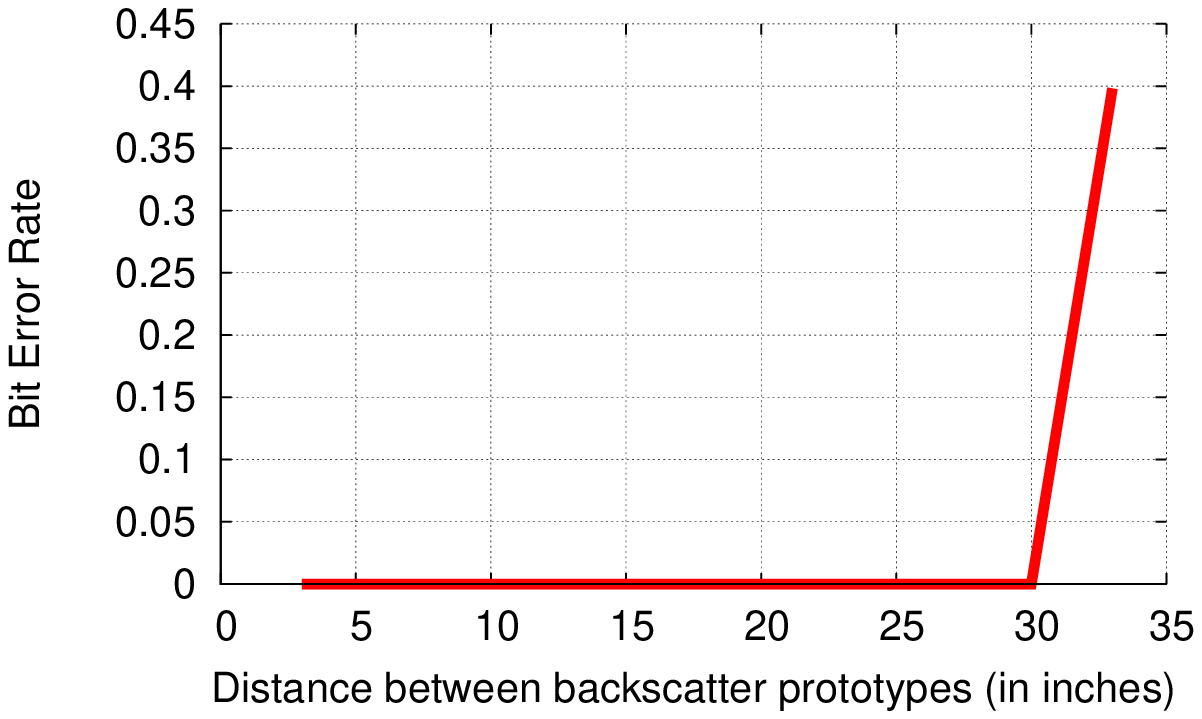}
        }
\vskip -0.15in
        \caption{{\bf BER for communication between two cards with integrated antennas.} Amazon card shown for comparison.}
\vskip -0.15in
\label{fig:card_2_card}
\end{figure}

We prototype proof-of-concept credit card form factor devices shown in Fig.~\ref{fig:card_2_card} that backscatter the single-tone from the Bluetooth device to communicate between each other. We replicate prior work on ambient backscatter~\cite{abc-sigcomm2013} and use its receiver architecture. We however, tune the hardware to operate at 2.4~GHz instead of the UHF TV band. The prototype also has an IGLOO Nano FPGA to implement digital baseband processing for \name\, an MSP430 micro-controller for running application software, LEDs for visual feedback, an accelerometer and capacitive touch sensors for user input.

We evaluate card to card communication by configuring one card to periodically transmit an 18 bit payload at 100~kbps to another, at time instances corresponding to the single-tone Bluetooth transmissions. The cards use their energy detectors to synchronize with the Bluetooth transmissions. We place the transmit card 3~inches from a 10~dBm TI Bluetooth device and vary the location of the receiver. Fig.~\ref{fig:card_2_card} shows the bit error rate achieved. The plot demonstrates that up to a distance of 30~inches, 10~dBm Bluetooth transmissions achievable on smartphones such as the Note 5 and iPhone 6, can enable card to card communication.

\section{Related Work}
Interscatter is related to our prior work on ambient and Wi-Fi backscatter communication~\cite{abc-sigcomm2013,abc-sigcomm2014, wifibackscatter-sigcomm2014} where devices communicate by backscattering ambient signals such as TV and Wi-Fi transmissions. Specifically, Wi-Fi backscatter~\cite{wifibackscatter-sigcomm2014} conveys information by either reflecting or not reflecting existing Wi-Fi packets on the wireless channel. This creates CSI changes that encode information. Since the minimum resolution at which the bits are conveyed is greater than a Wi-Fi packet, this limits the bit rate to 100--1000~bps. \textred{Recent work on FS-Backscatter~\cite{fs-backscatter} combines this packet-level backscatter approach with double sideband subcarrier modulation~\cite{nsdi16,rfid-textbook1,fm-backscatter-paper}. Specifically, it uses a ring oscillator to shift each Wi-Fi packet to a different frequency and achieves higher reliability than Wi-Fi backscatter. Interscatter differs from these designs in two ways. First, interscatter achieves standards-compliant 2--11~Mbps Wi-Fi packets where the content is generated by the backscattering device. This allows devices to transmit data at the high bit rates supported by Wi-Fi. In contrast, Wi-Fi backscatter and FS backscatter (in the context of Wi-Fi), encode information by either reflecting or not reflecting multiple packets from a Wi-Fi transmitter. As a result, the backscattered bits are encoded at the granularity of the Wi-Fi transmitter's packets and the backscatter device cannot change the packet content. Thus, the achieved bit rates are significantly lower. Second, these approaches require a large number of packets to be flooded by the Wi-Fi transmitter to send a small number of backscatter bits. This is not an efficient use of wireless resources and is likely to be unacceptable in crowded Wi-Fi networks. Finally in large scale production open loop ring oscillators are known to have frequency variation between devices~\cite{ring-otis, process-otis} and high phase noise which limit data rates and range performance. Given these constraints as well as the significant attenuation and detuning of antennas in human tissue, it is unclear whether this approach would be feasible for medical implants. }

\cite{nasa,wifibackscatter-sigcomm2015} use full-duplex radios to cancel the high-power Wi-Fi transmissions from the reader and decode weak backscattered signals to enable high data rates of 5--330 Mbps. These approaches require a custom full-duplex radio to be incorporated at the receiver and hence do not work with existing devices.


Passive Wi-Fi~\cite{nsdi16} generates 802.11b transmissions using double sideband subcarrier modulation~\cite{rfid-textbook1, fm-backscatter-paper} from a continuous wave transmitter. Specifically, it shows that backscatter can be used to generate 802.11b transmissions when backscattering transmissions from a custom transmitter that sends out constant wave signals.~\cite{bluetoothbackscatter} generates FSK Bluetooth transmissions by backscattering a continuous wave transmitter. We build on this work, but differ in three critical ways. 1) Prior work requires custom specialized continuous wave transmitter hardware. Thus, they cannot plug and play with commodity devices. We introduce a novel approach to backscatter communication where we demonstrate for the first time that one can transform transmissions from a Bluetooth device into Wi-Fi signals, and achieve a system that can work using only commodity radios. 2) Prior work could not perform single-sideband backscatter and hence used 44~MHz of bandwidth to generate 22~MHz Wi-Fi transmissions. This is undesirable given that the 2.4~GHz ISM band is becoming increasingly crowded. In contrast, we introduce the first single-sideband backscatter design that generates Wi-Fi signals on only one side of the carrier and has double the spectral efficiency over prior designs. 3) Prior work uses custom plugged-in hardware to communicate to the backscattering device. In contrast, we reuse Wi-Fi transmitters to convey information to the backscattering devices. Specifically, we introduce a novel mechanism that encodes information in OFDM symbols of a Wi-Fi transmission.

Finally, Anynfc~\cite{cheap_rfid_reader} sells RFID readers that connect to smartphones via the headphone jack for \$185. While promising, RFID readers do not have the same economy of scale as Wi-Fi or Bluetooth which cost a few dollars~\cite{ti_ble_chip_buy}. Further, existing devices already have Wi-Fi and Bluetooth radios. This takes us closer to the vision of backscatter as a general-purpose communication mechanism.

\section{Discussion and Conclusion}
We introduce a novel approach that transforms wireless transmissions from one technology to the other, on the air. Specifically, we show for the first time that Bluetooth transmissions can be used to generate Wi-Fi and ZigBee-compatible signals using backscatter communication. Using this approach, we build proof-of-concepts for previously infeasible applications including the first contact lens form-factor antenna prototype and an implantable neural recording interface that communicate directly with commodity devices such as smartphones and watches, thus enabling the vision of Internet connected implanted devices.

\textred{Finally we outline research directions for improving upon this work. First, interscatter enables high bit rates which allow data to be transmitted in shorter periods of time. This reduces the time for which transmissions occupy the channel and can reduce power consumption by duty cycling the device. Future work could improve overall throughput by using BLE data packets which can be transmitted at a faster rate than BLE advertisements. Additionally the latest Bluetooth standard increases the maximum length for these data packets which would further improve throughput. Lastly, while our current design focuses on generating 802.11b transmissions using backscatter, future work could explore OFDM based protocols such as 802.11g/n/ac to improve the bit rates we achieve by an additional order of magnitude.}


\bibliographystyle{abbrv}
\bibliography{paper}

\begin{thebibliography}{10}

\bibitem{altera_de1}
Altera de1 fpga development board.
\newblock \url{http://www.terasic.com.tw/cgi-bin/page/archive.pl?No=83}.

\bibitem{atheros_ar6003}
Atheros targets cellphone with wi-fi chip.
\newblock \url{http://www.eetimes.com/document.asp?doc_id=1172134}.

\bibitem{cadence_RFspectre}
Cadence rfspectre.
\newblock
  \url{http://www.cadence.com/products/rf/spectre_rf_simulation/pages/default.aspx}.

\bibitem{fcc-oneplus2}
Fcc id value: 2abz2-a0001.
\newblock \url{https://www.fcc.gov/general/fcc-id-search-page}.

\bibitem{fcc-samsungs6}
Fcc id value: A3lsmg920f.
\newblock \url{https://www.fcc.gov/general/fcc-id-search-page}.

\bibitem{fcc-samsungnote5}
Fcc id value: A3lsmn920i.
\newblock \url{https://www.fcc.gov/general/fcc-id-search-page}.

\bibitem{fcc-iphone6}
Fcc id value: Bcg-e2817a.
\newblock \url{https://www.fcc.gov/general/fcc-id-search-page}.

\bibitem{hmc190bms8}
Hms190bms8 by hittite microwave devices.
\newblock
  \url{https://www.hittite.com/content/documents/data_sheet/hmc190bms8.pdf}.

\bibitem{nasa}
Nasa news release: A wi-fi reflector chip to speed up wearables.
\newblock \url{http://www.jpl.nasa.gov/news/news.php?feature=4663}.

\bibitem{cheap_rfid_reader}
Rfid uhf portable mini reader writer.
\newblock \url{http://www.amazon.com/gp/product/B00W9IUALK}.

\bibitem{synopsis}
Synopsis design complier.
\newblock
  \url{http://www.synopsys.com/Tools/Implementation/RTLSynthesis/DesignCompiler/Pages/default.aspx}.

\bibitem{ti_ble_chip_buy}
Ti cc2650.
\newblock
  \url{http://www.digikey.com/product-detail/en/CC2650F128RHBR/CC2650F128RHBR-ND/5189550}.

\bibitem{rfid-textbook1}
Epc radio-frequency identity protocols class-1 generation-2 uhf rfid protocol
  for communications at 860 mhz-960mhz version 1.2.0, 2008.
\newblock
  \url{http://www.gs1.org/sites/default/files/docs/epc/uhfc1g2_1_2_0-standard-20080511.pdf}.

\bibitem{ecog-parkinson}
Man, machine and in between.
\newblock {\em Nature}, 2009.

\bibitem{ecog-bmi}
A 100-channel hermetically sealed implantable device for wireless neurosensing
  applications.
\newblock In {\em Circuits and Systems (ISCAS), 2012 IEEE International
  Symposium on}, May 2012.

\bibitem{lens3}
A.~J. Bandodkar and J.~Wang.
\newblock Non-invasive wearable electrochemical sensors: a review.
\newblock {\em Trends in Biotechnology}, 32(7):363 -- 371, 2014.

\bibitem{bmi1}
W.~Biederman, D.~Yeager, N.~Narevsky, A.~Koralek, J.~Carmena, E.~Alon, and
  J.~Rabaey.
\newblock A fully-integrated, miniaturized 10.5 uw wireless neural sensor.
\newblock {\em Solid-State Circuits, IEEE Journal of}, 2013.

\bibitem{usrp-802.11g}
B.~Bloessl, M.~Segata, C.~Sommer, and F.~Dressler.
\newblock {An IEEE 802.11a/g/p OFDM Receiver for GNU Radio}.
\newblock In {\em ACM SIGCOMM 2013, 2nd ACM SIGCOMM Workshop of Software Radio
  Implementation Forum (SRIF 2013)}, pages 9--16, Hong Kong, China, August
  2013. ACM.

\bibitem{scrambler-seed}
B.~Bloessl, C.~Sommer, F.~Dressier, and D.~Eckhoff.
\newblock The scrambler attack: A robust physical layer attack on location
  privacy in vehicular networks.
\newblock In {\em Computing, Networking and Communications (ICNC), 2015
  International Conference on}, pages 395--400, Feb 2015.

\bibitem{bluetoothbackscatter}
J.~Ensworth and M.~Reynolds.
\newblock Every smart phone is a backscatter reader: Modulated backscatter
  compatibility with bluetooth 4.0 low energy (ble) devices.
\newblock In {\em RFID, 2015 IEEE International Conference on}.

\bibitem{gabriel1996dielectric}
C.~Gabriel, S.~Gabriel, and E.~Corthout.
\newblock The dielectric properties of biological tissues: I. literature
  survey.
\newblock {\em Physics in medicine and biology}, 41(11):2231, 1996.

\bibitem{tep}
S.~Gollakota, N.~Ahmed, N.~Zeldovich, and D.~Katabi.
\newblock Secure in-band wireless pairing.
\newblock In {\em Proceedings of the 20th USENIX Conference on Security},
  SEC'11, pages 16--16, Berkeley, CA, USA, 2011. USENIX Association.

\bibitem{marnello}
B.~Han, A.~Schulman, F.~Gringoli, N.~Spring, B.~Bhattacharjee, L.~Nava, L.~Ji,
  S.~Lee, and R.~Miller.
\newblock Maranello: Practical partial packet recovery for 802.11.
\newblock In {\em Proceedings of the 7th USENIX Conference on Networked Systems
  Design and Implementation}, NSDI'10, pages 14--14, Berkeley, CA, USA, 2010.
  USENIX Association.

\bibitem{haykin2008communication}
S.~Haykin.
\newblock {\em Communication systems}.
\newblock John Wiley \& Sons, 2008.

\bibitem{hochberg2006neuronal}
L.~R. Hochberg, M.~D. Serruya, G.~M. Friehs, J.~A. Mukand, M.~Saleh, A.~H.
  Caplan, A.~Branner, D.~Chen, R.~D. Penn, and J.~P. Donoghue.
\newblock Neuronal ensemble control of prosthetic devices by a human with
  tetraplegia.
\newblock {\em Nature}, 442(7099):164--171, 2006.

\bibitem{wifibackscatter-sigcomm2014}
B.~Kellogg, A.~Parks, S.~Gollakota, J.~R. Smith, and D.~Wetherall.
\newblock Wi-fi backscatter: Internet connectivity for rf-powered devices.
\newblock In {\em Proceedings of the 2014 ACM Conference on SIGCOMM}, 2014.

\bibitem{allsee-nsdi14}
B.~Kellogg, V.~Talla, and S.~Gollakota.
\newblock Bringing gesture recognition to all devices.
\newblock In {\em Usenix NSDI}, volume~14, 2014.

\bibitem{nsdi16}
B.~Kellogg, V.~Talla, S.~Gollakota, and J.~Smith.
\newblock Passive wi-fi: Bringing low power to wi-fi transmissions.
\newblock In {\em Usenix NSDI}, 2016.

\bibitem{fm-backscatter-paper}
J.~Kimionis, A.~Bletsas, and J.~N. Sahalos.
\newblock Bistatic backscatter radio for power-limited sensor networks.
\newblock In {\em 2013 IEEE Global Communications Conference (GLOBECOM)}, pages
  353--358, Dec 2013.

\bibitem{koomey}
J.~Koomey, S.~Berard, M.~Sanchez, and H.~Wong.
\newblock Implications of historical trends in the electrical efficiency of
  computing.
\newblock {\em Annals of the History of Computing, IEEE}, 2011.

\bibitem{leuthardt2004brain}
E.~C. Leuthardt, G.~Schalk, J.~R. Wolpaw, J.~G. Ojemann, and D.~W. Moran.
\newblock A brain--computer interface using electrocorticographic signals in
  humans.
\newblock {\em Journal of neural engineering}, 1(2):63, 2004.

\bibitem{lens2}
Y.-T. Liao, H.~Yao, A.~Lingley, B.~Parviz, and B.~Otis.
\newblock A 3-uw cmos glucose sensor for wireless contact-lens tear glucose
  monitoring.
\newblock {\em Solid-State Circuits, IEEE Journal of}, 47(1):335--344, 2012.

\bibitem{liao20123}
Y.-T. Liao, H.~Yao, A.~Lingley, B.~Parviz, and B.~P. Otis.
\newblock A 3-cmos glucose sensor for wireless contact-lens tear glucose
  monitoring.
\newblock {\em Solid-State Circuits, IEEE Journal of}, 47(1):335--344, 2012.

\bibitem{abc-sigcomm2013}
V.~Liu, A.~Parks, V.~Talla, S.~Gollakota, D.~Wetherall, and J.~R. Smith.
\newblock Ambient backscatter: Wireless communication out of thin air.
\newblock In {\em Proceedings of the ACM SIGCOMM 2013 Conference on SIGCOMM},
  2013.

\bibitem{fullduplex-mobicom2014}
V.~Liu, V.~Talla, and S.~Gollakota.
\newblock Enabling instantaneous feedback with full-duplex backscatter.
\newblock In {\em Proceedings of the 20th Annual International Conference on
  Mobile Computing and Networking}, MobiCom '14, pages 67--78, New York, NY,
  USA, 2014. ACM.

\bibitem{ecog-limb}
D.~A. Lobel and K.~H. Lee.
\newblock Brain machine interface and limb reanimation technologies: Restoring
  function after spinal cord injury through development of a bypass system.
\newblock {\em Mayo Clinic Proceedings}, 89(5):708 -- 714, 2014.

\bibitem{mrd}
A.~Miu, H.~Balakrishnan, and C.~E. Koksal.
\newblock Improving loss resilience with multi-radio diversity in wireless
  networks.
\newblock In {\em MobiCom '05: Proceedings of the 11th annual international
  conference on Mobile computing and networking}, 2005.

\bibitem{bmi2}
R.~Muller, H.-P. Le, W.~Li, P.~Ledochowitsch, S.~Gambini, T.~Bjorninen,
  A.~Koralek, J.~Carmena, M.~Maharbiz, E.~Alon, and J.~Rabaey.
\newblock A minimally invasive 64-channel wireless ecog implant.
\newblock {\em Solid-State Circuits, IEEE Journal of}, 2015.

\bibitem{lens1}
J.~Pandey, Y.-T. Liao, A.~Lingley, R.~Mirjalili, B.~Parviz, and B.~Otis.
\newblock A fully integrated rf-powered contact lens with a single element
  display.
\newblock {\em Biomedical Circuits and Systems, IEEE Transactions on}, 2010.

\bibitem{abc-sigcomm2014}
A.~N. Parks, A.~Liu, S.~Gollakota, and J.~R. Smith.
\newblock Turbocharging ambient backscatter communication.
\newblock In {\em Proceedings of the 2014 ACM Conference on SIGCOMM}, 2014.

\bibitem{wifibackscatter-sigcomm2015}
D.~Pharadia, K.~R. Joshi, M.~Kotaru, and S.~Katti.
\newblock Backfi: High throughput wifi backscatter.
\newblock In {\em Proceedings of the 2015 ACM Conference on Special Interest
  Group on Data Communication}, 2015.

\bibitem{razavi1998rf}
B.~Razavi.
\newblock {\em RF microelectronics}, volume~1.
\newblock Prentice Hall New Jersey, 1998.

\bibitem{process-otis}
Y.~Su, J.~Holleman, and B.~P. Otis.
\newblock A digital 1.6 pj/bit chip identification circuit using process
  variations.
\newblock {\em IEEE Journal of Solid-State Circuits}, 43(1):69--77, Jan 2008.

\bibitem{bmi3}
V.~Talla, V.~Ranganathan, B.~Mahoney, and J.~Smith.
\newblock Dual band wireless power and bi-directional data link for implanted
  devices in 65 nm cmos.
\newblock ISCAS 2016.

\bibitem{matt-qambacksactter}
S.~Thomas and M.~Reynolds.
\newblock A 96 mbit/sec, 15.5 pj/bit 16-qam modulator for uhf backscatter
  communication.
\newblock In {\em RFID (RFID), 2012 IEEE International Conference on}, pages
  185--190, April 2012.

\bibitem{wyler1984subdural}
A.~R. Wyler, G.~A. Ojemann, E.~Lettich, and A.~A. Ward~Jr.
\newblock Subdural strip electrodes for localizing epileptogenic foci.
\newblock {\em Journal of neurosurgery}, 60(6):1195--1200, 1984.

\bibitem{yao2012contact}
H.~Yao, Y.~Liao, A.~Lingley, A.~Afanasiev, I.~L{\"a}hdesm{\"a}ki, B.~Otis, and
  B.~Parviz.
\newblock A contact lens with integrated telecommunication circuit and sensors
  for wireless and continuous tear glucose monitoring.
\newblock {\em Journal of Micromechanics and Microengineering}, 22(7):075007,
  2012.

\bibitem{lens4}
H.~Yao, Y.~Liao, A.~R. Lingley, A.~Afanasiev, I.~Lähdesmäki, B.~P. Otis, and
  B.~A. Parviz.
\newblock A contact lens with integrated telecommunication circuit and sensors
  for wireless and continuous tear glucose monitoring.
\newblock {\em Journal of Micromechanics and Microengineering}, 22(7):075007,
  2012.

\bibitem{ring-otis}
D.~Yeager, F.~Zhang, A.~Zarrasvand, N.~T. George, T.~Daniel, and B.~P. Otis.
\newblock A 9ua, addressable gen2 sensor tag for biosignal acquisition.
\newblock {\em IEEE Journal of Solid-State Circuits}, 45(10):2198--2209, Oct
  2010.

\bibitem{fs-backscatter}
P.~Zhang, M.~Rostami, P.~Hu, and D.~Ganesan.
\newblock Enabling practical backscatter communication for on-body sensors.
\newblock In {\em Proceedings of the ACM SIGCOMM 2016 Conference on SIGCOMM},
  2016.

\end{thebibliography}

%
%
\end{sloppypar}
\label{lastpage}
\end{document}